\documentclass[useAMS,usenatbib]{mn2e}

\usepackage{xspace}
\usepackage{graphicx}
\usepackage{color}
\def\nsample{160}

\def\growth{$1.8 \pm 0.3$}
\def\growthunmatched{$1.55 \pm 0.18$}
\def\growthadj{2.2}
\def\fitindex{$1.6 \pm 0.2$}

\title[Brightest Cluster Galaxies]{Evidence for Significant Growth in
  the Stellar Mass of 
Brightest Cluster Galaxies over the Past 10 Billion Years.}
\author[C. Lidman et al.]
  {C. Lidman,$^1$\thanks{E-mail: clidman@aao.gov.au}
    J. Suherli,$^{1,2}$ A. Muzzin,$^3$ G. Wilson,$^4$ R. Demarco,$^5$ S. Brough,$^1$ 
     \newauthor
    A. Rettura,$^4$, J. Cox,$^4$ A. DeGroot,$^4$ H. K. C. Yee,$^6$ D. Gilbank,$^7$ H. Hoekstra,$^3$
     \newauthor
   M. Balogh,$^8$ E. Ellingson,$^9$ A. Hicks,$^{10}$  J. Nantais,$^5$ A. Noble,$^{11}$ M. Lacy,$^{12}$
    J. Surace,$^{13}$
     \newauthor
   T. Webb$^{10}$\\
  $^1$Australian Astronomical Observatory, PO Box 296, Epping NSW 1710, Australia\\
  $^2$Bosscha Observatory, Institut Teknologi Bandung, Lembang, Bandung, West Java, Indonesia\\
  $^3$Leiden Observatory, Leiden University, PO Box 9513, 2300 RA Leiden, The Netherlands\\
  $^4$Department of Physics and Astronomy, University of California, Riverside, CA 92521\\
  $^5$Department of Astronomy, Universidad de Concepcion, Casilla 160-C, Concepcion, Chile\\
  $^6$Department. of Astronomy \& Astrophysics, University of
  Toronto, 50 St. George St., Toronto, Ontario, Canada, M5S 3H4 TBA\\
  $^7$South African Astronomical Observatory, PO Box 9, Observatory,
  7935, South Africa\\
  $^8$Department of Physics and Astronomy, University of Waterloo,
  Waterloo, Ontario N2L 3G1, Canada\\
  $^9$Center for Astrophysics and Space Astronomy, 389UCB, University
  of Colorado, Boulder, CO 80309, USA\\
  $^{10}$Department of Physics and Astronomy, Michigan State University,
  East Lansing, MI 48824-2320, USA\\
  $^{11}$Department of Physics, McGill University, Montréal, QC,
  Canada\\
  $^{12}$North American ALMA Science Center, NRAO Headquarters, 520
  Edgemont Road, Charlottesville, VA 22903\\
  $^{13}$Spitzer Science Center, California Institute
 of Technology, 220-6, Pasadena, CA, 91125\\
}

\begin{document}

\date{Accepted YYYY Month DD. Received YYYY Month DD}

\pagerange{\pageref{firstpage}--\pageref{lastpage}} \pubyear{2002}

\maketitle

\label{firstpage}

\begin{abstract} Using new and published data, we construct a sample
  of \nsample\ brightest cluster galaxies (BCGs) spanning the redshift
  interval $0.03 < z < 1.63$. We use this sample, which covers 70\% of
  the history of the universe, to measure the growth in the stellar
  mass of BCGs after correcting for the correlation between the
  stellar mass of the BCG and the mass of the cluster in which it
  lives. We find that the stellar mass of BCGs increase by a factor of
  \growth\ between $z=0.9$ and $z=0.2$. Compared to earlier works, our
  result is closer to the predictions of semi-analytic
  models. However, BCGs at $z=0.9$, relative to BCGs at $z=0.2$, are
  still a factor of 1.5 more massive than the predictions of these
  models. Star formation rates in BCGs at $z\sim1$ are generally too
  low to result in significant amounts of mass. Instead, it is likely
  that most of the mass build up occurs through mainly dry mergers in
  which perhaps half of the mass is lost to the intra-cluster medium
  of the cluster.

\end{abstract}


\section{Introduction}

Brightest Cluster Galaxies (BCGs) are amongst the largest, most
luminous and most massive galaxies in the universe at the present
epoch. Located in the cores of rich galaxy clusters, BCGs are easy to
identify, both observationally and in simulations. They can
also be observed at a time when the universe was less than a third of
its current age. They therefore provide an attractive target for
testing our understanding of the processes that drive galaxy evolution,
albeit in the most massive galaxies of the universe.

In the hierarchical scenario for the formation of structure in our
universe, galaxies start off as small fluctuations in the density of
matter and build up their stellar mass over time by converting
material accreted from their surroundings into stars and by merging
with other galaxies \citep[see][for a review]{Baugh2006}. 
In semi-analytic models that use the hierarchical scenario as their
foundation, the stellar mass of a BCG increases significantly with
time. For example, between redshift $z=1.0$ (corresponding to a look-back
time of 6.7\,Gyr) to $z=0$, the semi-analytic model described in
\citet{DeLucia2007} predicts that BCGs increase their stellar mass by
a factor of four \citep{DeLucia2007}.

In contrast to this prediction, observations appear to suggest that
there is little growth in the stellar mass of BCGs, although
apparently conflicting results have been reported. Using a sample of
optically selected clusters, \citet{Aragon-Salamanca1998} found that
the stellar mass of BCGs grew by a factor of 4 between $z=1$ and
today. \citet{Burke2000}, on the other hand, using a sample of X-ray
selected clusters over a similar redshift range, find substantially
less growth.\footnote{Both \citet{Aragon-Salamanca1998} and
  \citet{Burke2000} use an Einstein de-Sitter universe,
  i.e.~$\Omega_{\mathrm M},\Omega_{\Lambda}=1,0$, with $H_0=50$
  km\,s$^{-1}$\,Mpc$^{-1}$ for the cosmology. While their results are
  not directly comparable to the results in later papers, one can
  compare the results of the two papers.}  \citet{Burke2000} conclude
that sample selection can explain part of the difference between their
results and those in \citet{Aragon-Salamanca1998}, a conclusion that
was supported by \citet{Nelson2002}. 
%The clusters in \citet{Burke2000}
%are bright in X-rays, whereas those in \citet{Aragon-Salamanca1998}
%were predominantly selected in the optical. 
%A difference in the average cluster mass between the two samples will
%lead to differences, since the mass of the BCG correlates with the
%mass of the cluster
%\citep{Edge1991,Burke2000,Brough2002,Brough2008,Stott2008}, although
%\citet{Whiley2008} noted that the correlation is quite weak. 
In an independent study, using an optically selected sample of 21
high-redshift clusters, \citet{Whiley2008} find little change in the
stellar mass of BCGs since $z\sim 1$.

At higher redshifts, the discrepancy between the models and the
observations is larger. \citet{Collins2009} and \citet{Stott2010}, using a sample
of 20 mostly X-ray selected clusters and a sample of nearby clusters
from \citet{Stott2008}, find that there is little growth between $z
\sim 1.4$ and now. At $z \sim 1.4$, the semi-analyic model of
\citet{DeLucia2007} predicts that BCGs should be a factor of six
less massive. Therefore, there appears to be
clear disagreement between the models and the observations.

In this paper, we expand upon the work that has been done so far in
three ways. First, we increase the number of BCGs beyond $z=0.8$ for
which accurate near-IR photometry is available. Second, we extend the
redshift baseline by including the BCGs in two recently discovered
clusters at $z\sim 1.6$. Third, we use our large sample to account for
the correlation between the stellar mass of the BCG and the mass of
the cluster in which it lives.

We start the paper with a description of our new sample of BCGs in Section
2, followed, in Section 3, with a description of the near-IR imaging
data that we use in later sections. In Sections 4 and 5, we derive the
magnitudes and colours of the BCGs in our sample and compare them to
predictions made by simple and composite stellar population
models.  Following \citet{Stott2010}, we use this comparison to
estimate stellar masses. In Section 6, we discuss our results,
comparing them to the predictions made by semi-analytic models and
examining how robust they are to our methods.  In the final section,
we summarise our main results.  Throughout the paper, all magnitudes
and colours are measured in the observer frame and are placed on the
2MASS photometric system. Vega magnitudes are used throughout the
paper. We also assume a flat $\Lambda$CDM cosmology with
$\Omega_{\Lambda}=0.73$ and $H_0=70$ km\,s$^{-1}$\,Mpc$^{-1}$. %\citep{Suzuki2012}.

\section{A New Sample of Distant BCGs}

We use clusters from the SpARCS\footnote{Spitzer Adaptation of the
  Red-Sequence Cluster Survey, www.faculty.ucr.edu/$\sim$gillianw/SpARCS/}
survey \citep{Muzzin2009a,Wilson2009a,Demarco2010a,Muzzin2012} to
assemble a sample of 12 BCGs spanning the redshift interval $0.85 < z
< 1.63$. The coordinates and redshifts of the clusters are listed in
Table~\ref{tab:Observations}. Ten of the twelve clusters were observed
in the GCLASS\footnote{Gemini Cluster Astrophysics
  Spectroscopic Survey, www.faculty.ucr.edu/$\sim$gillianw/GCLASS/} survey,
which used the Gemini Multi-object Spectrographs on Gemini North and
Gemini South Telescopes to obtain between 20 and 80
spectroscopically confirmed members per cluster
\citep{Muzzin2012}. The other two clusters, which are the most distant
clusters in our sample, are more recent discoveries. Both clusters are
spectroscopically confirmed, with a dozen spectroscopic redshifts per
cluster (Muzzin et al. in preparation; Wilson et al. in preparation).

All twelve clusters were discovered by searching for over-densities in
the number of red galaxies using a combination of images taken with
IRAC on the Spitzer Space Telescope with z-band images taken with
either MegaCam on the Canada-France-Hawaii Telescope (CFHT) or MOSAIC II on the Cerro Tololo Blanco
Telescope. The ten GCLASS clusters were found using the
z-[3.6] colour, whereas the two more distant clusters were found
using the [3.6]-[4.5] colour together with the requirement of a red
z-[3.6] colour.  Further details of how the clusters were discovered can be found in
\citep{Muzzin2008a,Muzzin2009a,Wilson2009a} and (Muzzin et al., in
preparation). 

\section{Observations}

We used three near-IR imaging cameras to observe 12 clusters.  Six of
the clusters were observed with the Wide-field InfraRed Camera
(WIRCam) on the Canada-France-Hawaii Telescope (CFHT) on Mauna
Kea. Another three clusters were imaged with the Infrared Side Port
Imager (ISPI) on the Blanco Telescope at the Cerro Tololo
Inter-American Observatory (CTIO). Finally, the three most distant
clusters were imaged with the High Acuity Wide field K-band Imager
(HAWK-I) on Yepun (VLT-UT4) at the ESO Cerro Paranal Observatory. The
fields-of-view of the imagers, and their plate scales are noted
in Table~\ref{tab:Instruments}.  Details of the observations,
including exposure times, are listed in
Table~\ref{tab:Observations}. With the exception of the two most
distant clusters (SpARCS~J033056-284300
and SpARCS~J022426-032331), all clusters were imaged in J and Ks. At
the time SpARCS~J033056-284300
and SpARCS~J022426-032331 were observed with HAWK-I, neither cluster
had been spectroscopically confirmed.  The Y and Ks filter pair were chosen over
J and Ks, since the former pair almost straddle the 4000\,$\AA$
break, thereby increasing the contrast of cluster members over
field galaxies and easing target selection for spectroscopy.

\begin{table*}
\caption{Observational summary}\label{tab:Observations}
\label{table:summary}  
\centering           
\begin{tabular}{l r r r l r r}
\hline
Cluster & Redshift & R.A.$^5$  & Decl. $^5$ & Instrument/Telescope &\multicolumn{2}{c}{Exposure times: J and Ks}\\
         &          & J2000     & J2000      &                     &  [s] & [s] \\
\hline
SpARCS J003442-430752\,$^1$     & 0.867 & 00:34:42.03 & -43:07:53.4  & ISPI/Blanco            & 17280 & 8800 \\
SpARCS J003645-441050\,$^1$     & 0.869 & 00:36:44.99 & -44:10:49.8  & ISPI/Blanco            & 17280 & 7080 \\
SpARCS J161314+564930\,$^{1,2}$  & 0.871 & 16:13:14.63 &  56:49:30.0  & WIRCAM/CFHT            & 6240  & 6300 \\
SpARCS J104737+574137\,$^1$     & 0.956 & 10:47:33.43 &  57:41:13.4  & WIRCAM/CFHT            & 7560  & 2400 \\
SpARCS J021524-034331\,$^1$     & 1.004 & 02:15:23.99 & -03:43:32.2  & ISPI/Blanco            & 26640 & 11800 \\
SpARCS J105111+581803\,$^1$     & 1.035 & 10:51:11.22 &  58:18:03.3  & WIRCAM/CFHT            &  6840 & 2700 \\
SpARCS J161641+554513\,$^{1,2}$  & 1.156 & 16:16:41.32 &  55:45:12.4  & WIRCAM/CFHT            & 18960 & 7000 \\
SpARCS J163435+402151\,$^{1,3}$  & 1.177 & 16:34:38.21 & 40:20:58.4  & WIRCAM/CFHT            & 11640 & 6850 \\
SpARCS J163852+403843\,$^{1,3}$  & 1.196 & 16:38:51.64 &  40:38:42.8  & WIRCAM/CFHT            & 11640 & 6000 \\
SpARCS J003550-431224\,$^{1,4}$  & 1.335 & 00:35:49.68 & -43:12:23.8 & HAWK-I/Yepun  & 11040 & 12000 \\
\hline 
Cluster & Redshift & R.A.$^5$  & Decl. $^5$ & Instrument/Telescope &\multicolumn{2}{c}{Exposure times: Y and Ks}\\
         &          & J2000     & J2000      &                     &  [s] & [s] \\
\hline
SpARCS J033056-284300          & 1.626 & 03:30:55.87 &-28:42:59.7  & HAWK-I/Yepun   & 8880 & 3040 \\
SpARCS J022426-032331    & 1.633 & 02:24:26.32 & -03:23:30.7 & HAWK-I/Yepun  & 8640 & 5040 \\

\hline
\multicolumn{7}{l}{Note 1: \citet{Muzzin2012}} \\
\multicolumn{7}{l}{Note 2: \citet{Demarco2010a}}\\
\multicolumn{7}{l}{Note 3: \citet{Muzzin2009a}} \\
\multicolumn{7}{l}{Note 4: \citet{Wilson2009a}} \\
\multicolumn{7}{l}{Note 5: Coordinates of the BCG} \\  % Used the NIR K band images
%\multicolumn{6}{l}{Note 6: Observed in Y and Ks} \\ 

\end{tabular}
\end{table*}

% Observing dates, if needed
%SpARCS J003442-430752 2009-10-26 to 2009-10-31 
%SpARCS J003645-441050 2009-10-27 to 2009-10-29 
%SpARCS J161314+564930 2009-08-04 to 2010-09-25 
%SpARCS J104737+574137 2010-03-28 to 2010-04-03 
%SpARCS J021524-034331 2009-10-26 to 2009-10-31 
%SpARCS J105111+581803 2010-03-25 to 2010-04-01 
%SpARCS J161641+554513 2009-06-13 to 2010-09-27 
%SpARCS J163435+402151 2009-08-05 to 2010-09-27 
%SpARCS J163852+403843 2010-04-30 to 2010-05-07 
%SpARCS J003550-431224 2009-10-06 to 2009-11-03 
%SpARCS J033056-284300 2010-07-19 to 2010-08-27 
%SpARCS J022426-032331 2010-07-25 to 2010-08-16 

\begin{table*}
\caption{Instrument summary}\label{tab:Instruments}
\label{table:summary}  
\centering           
\begin{tabular}{l l r r l}
\hline 
Instrument & Telescope & Pixel Scale & FoV  & Detector\\
& & [\arcsec] & [\arcmin] & \\
\hline 
WIRCAM${^1}$ & CFHT & 0.304 & 20.5 & 2x2 Hawaii-2RG mosaic\\
ISPI${^2}$ & Blanco & 0.307& 10.3 & Hawaii-2 \\
HAWK-I${^3}$ & Yepun (VLT-UT4) & 0.1065& 7.5 & 2x2 Hawaii-2RG mosaic\\
\hline
\multicolumn{5}{l}{Note 1: \citep{Puget2004a}}\\
\multicolumn{5}{l}{Note 2: \citep{vanderBliek2004}} \\
\multicolumn{5}{l}{Note 3: \citep{Pirard2004a,Casali2006a}}\\
\end{tabular}
\end{table*}

\subsection{Data Reduction}

The processing of the raw data was done in a standard manner and
largely follows the steps outlined in \citet{Lidman2008a}.  Data from
each of the cameras were pre-processed (dark subtraction,
flat-fielding, and sky subtraction) using a combination of
observatory-developed instrument pipelines (for example, the CFHT data
were processed with version 1.0 of the `I`iwi pipeline\footnote{http://www.cfht.hawaii.edu/Instruments/Imaging/WIRCam/}) and
  our own scripts using IRAF\footnote{IRAF is distributed by the
    National Optical Astronomy Observatories which are operated by the
    Association of Universities for Research in Astronomy, Inc., under
    the cooperative agreement with the National Science Foundation}.

  SCAMP (version 1.6.2) and SWarp (version
  2.17.6)\footnote{http://www.astromatic.net/} were used to map the
  sky-subtracted images onto a common astrometric reference
  frame. After accounting for gain variations between chips (only
  relevant for the data that were taken with HAWK-I and WIRCAM) and creating
  individual bad pixel maps to account for bad pixels and remnants
  from bright stars observed in previous frames,
  the images were then combined with the {\tt imcombine} task within
  IRAF.  Each image was weighted with the inverse square of the FWHM
  of the PSF.

  With the exception of the data taken in the Y band, zero points were
  set using stars from the 2MASS point source catalogue
  \citep{Skrutskie2006a}. Typically, between 10 to 40 unsaturated
  2MASS stars with 2MASS quality flags of 'A' or 'B' were selected to
  measure zero points and their uncertainties.  2MASS stars were
  weighted by the reported uncertainties in the 2MASS point source
  catalogue. The uncertainties in the zero points are generally less
  than 2\%, and more typically 1\%, for both J and Ks. For Y, the zero
  point was set using standard stars that were observed during the
  same night as the clusters. The uncertainty is estimated from the
  night-to-night variation in the zero points and is around 2\%.

\subsection{Data quality}

Overall, the depth and quality of the imaging data varies
substantially from one image to another.  The image quality, as
measured from bright stars, varies from 0\farcs3 in the data taken
with HAWK-I to 1\farcs5 in the data taken with ISPI. 

The image depth, which we define as the 5 sigma point source detection
limit, varies from 19.5 for the Ks band image of SpARCS J003645-441050
to 25.1 for the Y band image of SpARCS J022426-032331. In all cases,
the BCG is at least 2 mag brighter than the detection limit. The
median signal-to-noise ratio is around
50. Table~\ref{table:datasummary} summarises the image quality and
image depth.

% A table summarizing the image quality, the depth, noting how it was computed.

\begin{table*}
\caption{Image quality and image depth}
\label{table:datasummary}  
\centering           
\begin{tabular}{l c c || c c}
\hline
Cluster & Image quality & Image depth$^1$ & Image quality & Image depth$^1$ \\
        & [\arcsec]     & [mag]           & [\arcsec]     & [mag]           \\
\hline
        & \multicolumn{2}{c}{J} & \multicolumn{2}{c}{Ks}\\
       
\hline
SpARCS J003442-430752  &  1.25 & 21.8  &  0.98  & 19.9 \\ % Cluster B
SpARCS J003645-441050  &  1.13 & 21.7  &  1.47  & 19.5 \\ % Cluster C
SpARCS J161314+564930  & 0.77 & 22.2  &  0.72  & 21.1 \\
SpARCS J104737+574137  & 0.69 & 22.2  &  0.60  & 21.2 \\
SpARCS J021524-034331  & 1.07 & 21.8  &  0.89  & 20.3 \\ % Cluster D
SpARCS J105111+581803  & 0.66 & 22.5  & 0.74  & 20.6 \\
SpARCS J161641+554513  & 0.70 & 22.8  & 0.75  & 21.2 \\
SpARCS J163435+402151  & 0.65 & 22.9  & 0.67  & 21.2 \\
SpARCS J163852+403843  & 0.61 & 23.1  & 0.58  & 21.5 \\
SpARCS J003550-431224   & 0.35 & 24.6  & 0.31  & 23.1 \\
\hline
        & \multicolumn{2}{c}{Y} & \multicolumn{2}{c}{Ks}\\
\hline
SpARCS J033056-284300  &   0.45 & 24.0 &  0.29 & 21.9\\       
SpARCS J022426-032331  &   0.34 & 25.1 &  0.51 & 21.5\\ 

% Differences seem too large between frames (esp. b/n the two Y band
%frames ...)
%Cannot see an error in the code.

\hline
\end{tabular}
\medskip
\begin{minipage}{156mm}
$^1$The image depth is the 5 sigma point-source detection limit measured over an aperture that has a diameter that is twice the image quality.
\end{minipage}
\end{table*}

\section{Analysis}

\begin{figure*}
\includegraphics[width=16cm]{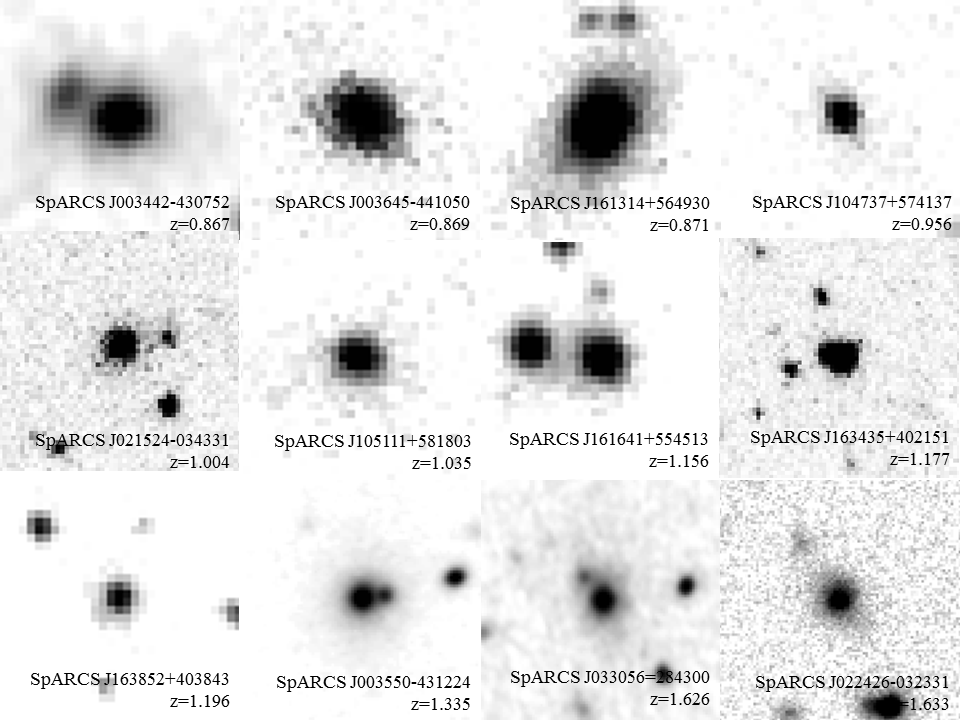}
\caption{ Ks-band cutouts of the 12 BCGs used in this paper. The
  images are 9\arcsec\ on a side, which corresponds to 70\,kpc for the
  nearest BCG and 78\,kpc for the most
  distant.}\label{fig:BCGimages}
\end{figure*}

Identifying the brightest galaxy\footnote{Throughout this paper, the
  BCG is defined as the brightest cluster member in the observer-frame
  Ks band.} in each cluster was generally straightforward. Images of
the BCGs are shown in Fig.~\ref{fig:BCGimages}, and their coordinates
are listed in Table~\ref{tab:Observations}.  With only two exceptions
-- SpARCS J105111+581803 and SpARCS J163435+402151 -- the BCGs are
located near to the projected centre of the clusters. For both SpARCS
J105111+581803 and SpARCS J163435+402151, the BCGs are $\sim250$\,kpc
from the projected centre of the cluster. The projected distances are
not excessively large when compared to low-redshift clusters
\citep{Bildfell2008,Sehgal2012}, and both BCGs have redshifts that place them
within 300 km/s of cluster redshift.

\subsection{Photometry}\label{sec:photom}

To estimate total magnitudes of the BCGs in the SpARCS clusters, we
follow \citet{Stott2010} and use the SExtractor\footnote{We used
  version 2.6.6 of SExtractor - http://www.astromatic.net/ - in double
  image mode.} {\tt MAG\_AUTO} magnitude.  {\tt MAG\_AUTO} is a
Kron-like magnitude \citep{Kron1980} within an elliptical
aperture. For a given object, the elongation and orientation of the
aperture is determined by second order moments of the light
distribution.  In this paper, the size of the aperture is set to the
standard value of 2.5 times the first raw moment. Note that the
definition of the first raw moment used by SExtractor differs from the one used in
\citep{Kron1980}. See the SExtractor user's manual and
\citet{Graham2005} for further details. Other SExtractror parameters are set to their
default values. For example, the background is determined globally and
the minimum Kron radius is set to 3.5, the units of which are not
pixels but in units of the semi-major (or semi-minor) axis.

For measuring colours, we first match the image quality between images
using the IRAF {\tt psfmatch} task and then measure the flux in apertures
that have a diameter of 16\,kpc. We use the same physical diameter for
all BCGs. At $z\sim 1$, this projects to
$\sim2\farcs1$ on the sky. The apertures we use are double the size of
the apertures used in \citet{Stott2010}. The image quality of the
poorest images - the Ks-band image of SpARCS J003645-441050, for
example - is not sufficiently good enough to use apertures this
small. 

We investigated how the colours change with the size of the
apertures. We varied the aperture diameter from 10\,kpc to 24\,kpc.
With three exceptions, the colours change by less than 2\%, which is
similar to the statistical
uncertainty. 
The exceptions are the BCGs in SpARCS J105111+581803 and our two most
distant clusters, where we see changes of up to 6\%.  Interestingly, the
BCG of SpARCS J105111+581803 is about 250\,kpc from the centre of the
cluster, is an [OII] emitter, and has, relative to other BCGs, a blue
colour. Our two most distant BCGs, which were observed in Y and Ks,
are also [OII] emitters. The change in colour with aperture diameter
might indicate that these galaxies have substantial colour
gradients. The other BCGs do not show any evidence for colour
gradients over the range of apertures explored.

Errors in the photometry are dominated by sky noise, so they were
estimated by examining the distribution of the integrated counts in
apertures that were randomly placed in regions that were free of
objects. For colours, the errors were estimated for each filter
separately and then added in quadrature.

The filter transmission curves of the J and Ks bands in ISPI, WIRCam
and HAWK-I are similar to one another; however, they differ slightly
from the filter curves of the respective filters used in 2MASS. To
account for this difference we offset the colours by an amount that
depends on i) the average spectral energy distribution (SED) of the
stars used to determine the image zero points, ii) the SED of the
BCG and iii) its redshift. To determine the offset, we assume that the
average star can be modelled as a K5 dwarf, which has a J-Ks colour that
is similar to the average colour of stars that are used to determine
the zero-point, and that the SED of the BCG corresponds to the one
predicted by model \#3 in Fig.~\ref{fig:models}. See Sec.~\ref{sec:colour} for a
detailed description of this model. Between redshift 0.8 and 1.6, the magnitude of the
correction is about 0.08 mag. At these redshifts, the dependence of
the correction on redshift and the assumed spectrum of the BCG is
slight, with extreme values differing by 0.03 mag.

The magnitudes and colours of the 12 BCGs in our sample 
are shown in Table~\ref{tab:results}. 

\begin{table*}
\caption{SpARCS BCG photometry}\label{tab:results}
\centering           
\begin{tabular}{lrccrr}
\hline
Cluster                   & Redshift & Ks    & J-Ks \\
                         &          & [mag] & [mag] \\
\hline
SpARCS J003442-430752 & 0.867 & 16.516 (0.039) & 1.863 (0.033)\\
SpARCS J003645-441050 & 0.867 & 16.092 (0.047) & 1.837 (0.030)\\
SpARCS J161314+564930 & 0.873 & 15.693 (0.015) & 1.794 (0.012)\\
SpARCS J104737+574137 & 0.956 & 17.140 (0.031) & 1.889 (0.029)\\
SpARCS J021524-034331 & 1.004 & 16.876 (0.140) & 1.861 (0.052)\\
SpARCS J105111+581803 & 1.035 & 16.877 (0.046) & 1.740 (0.030)\\
SpARCS J161641+554513 & 1.156 & 17.017 (0.031) & 1.729 (0.022)\\
SpARCS J163435+402151 & 1.177 & 17.349 (0.023) & 1.839 (0.026)\\
SpARCS J163852+403843 & 1.196 & 17.647 (0.052) & 1.913 (0.051)\\
SpARCS J003550-431224 & 1.340 & 17.524 (0.014) & 1.981 (0.009)\\
SpARCS J033056-284300 & 1.620 & 17.881 (0.041) & ... (...) \\
SpARCS J022426-032331 & 1.630 & 18.071 (0.026) & ... (...) \\
 
\hline
\end{tabular}
\end{table*}

\subsection{External samples}\label{sec:samples}

To our sample of 12 high-redshift BCGs, we add BCGs from a number of external
samples.  At low to intermediate redshifts ($z=0.04$ to $z=0.83$), we
add 103 of the 104 BCGs from \citet{Stott2008}, excluding the BCG of
MS1054.5-0321. At higher redshifts ($z=0.81$ to $z=1.46$), we use a
sample of 20 BCGs from \citet{Stott2010}. The BCG of MS1054.5-0321 is
common to both samples. Like the 12 BCGs in our sample, the BCGs from
these two external samples were observed in J and Ks.

For the $z<0.15$ BCGs in \citet{Stott2008}, \citet{Stott2008} used
photometry from the extended and point source catalogues of 2MASS
\citep{Skrutskie2006a}.  For the $z> 0.15$ BCGs, \citet{Stott2008}
used the {\sf SExtractor} {\tt MAG\_BEST} magnitude. Depending on the
level of crowding, {\tt MAG\_BEST} is either a corrected isophotal
magnitude {\tt MAG\_ISOCOR} or the {\tt MAG\_AUTO} magnitude
\citep{Bertin1996}.

The BCGs in the high-redshift sample of \citet{Stott2010} are hosted
by clusters that come from a number of sources. Not all of the
clusters are X-ray selected; however, all are X-ray luminous, with
X-ray luminosities exceeding $10^{44}$\,erg/s. The photometry of these
BCGs is measured with the {\sf SExtractor} {\tt MAG\_AUTO} magnitude.
The BCGs in the low-to-intermediate-redshift sample of
\citet{Stott2008} are all hosted by clusters that have X-ray
luminosities in excess of $10^{44}$\,erg/s (in the 0.1--2.4 KeV band).

Additional BCG samples have been published in the
literature. \citet{Aragon-Salamanca1998} published k-corrected K-band
magnitudes for BCGs in 25 clusters up to $z=0.92$. \citet{Whiley2008}
combined this sample with 2MASS photometry of the low-redshift BCG
sample of \citet{vonderLinden2007} and their own photometry of a
sample of 21 intermediate-to-high redshift ($0.39 < z < 0.96$) BCGs
from the ESO Distant Cluster Survey. The photometry of all these
samples are measured in fixed $37$\,kpc diameter apertures, and is
converted to the rest-frame K-band using k-corrections. In this paper,
as in the papers of \citet{Stott2008} and \citet{Stott2010}, we do not
apply k-corrections and we measure the flux in differently sized
apertures. These differences mean that we cannot use the photometry
from these studies directly without inverting the k-corrections and
applying a correction for the different size of the apertures. Without
reanalysing the data, the latter is difficult to estimate, so we do
not add the BCGs from these samples to ours.

In addition to the BCGs in \citet{Stott2010}, we add BCGs in 15 X-ray luminous
clusters from the intermediate redshift CNOC1\footnote{Canadian
  Network for Observational Cosmology} cluster sample
\citep{Yee1996}. The CNOC1 clusters are from the Einstein Medium
Sensitivity Survey \citep{Gioia1990}. We use the Ks-band photometry
from Muzzin et al. (2007a); we note that these clusters were not
observed in the J band.

Ks-band imaging data for 14 of the 15 CNOC1 clusters were
obtained using the Ohio State-NOAO Infrared Imaging Spectrograph
(ONIS) on the Kitt Peak National Observatory (KPNO) 2.1 m telescope.
ONIS has a pixel scale of 0.288\arcsec, which is similar to the pixel
scale of the cameras used to observe most of the clusters in our
SpARCS sample.  One cluster, MS~0440+02 was obtained using the PISCES
camera on the Steward Observatory 90 inch (2.3 m) telescope. PISCES
has a pixel scale of 0.495\arcsec. The image quality in the fully
reduced images, varies between 0.7 and 1.3\arcsec\
\citep[See][for further details]{Muzzin2007a}.  

We have reanalysed the processed Ks-band images of clusters in the
CNOC1 sample following the procedure used for clusters in our SpARCS sample
(see Section~\ref{sec:photom}). The Ks-band magnitude of these galaxies
is reported in Table~\ref{tab:CNOC1}. 

\begin{table}
\caption{CNOC1 BCG photometry}\label{tab:CNOC1}
\centering           
\begin{tabular}{lrccc}
\hline
Cluster                   & Redshift & Ks    \\
                         &          & [mag] \\
\hline
 A2390                          & 0.228 & 13.489 (0.068)\\
 MS0440+02                      & 0.197 & 13.337 (0.054)\\
 MS0451+02                      & 0.201 & 13.938 (0.065)\\
 MS0839+29                      & 0.193 & 13.411 (0.062)\\
 MS1006+12                      & 0.261 & 13.786 (0.074)\\
 MS1231+15                      & 0.235 & 13.891 (0.065)\\
 MS1455+22                      & 0.257 & 13.558 (0.062)\\
 MS0016+16                      & 0.547 & 15.288 (0.078)\\
 MS0302+16                      & 0.425 & 15.008 (0.065)\\
 MS0451-03                      & 0.539 & 15.176 (0.071)\\
 MS1008-12                      & 0.306 & 13.676 (0.076)\\
 MS1224+20                      & 0.326 & 14.409 (0.078)\\
 MS1358+62                      & 0.329 & 14.292 (0.063)\\
 MS1512+36                      & 0.373 & 14.632 (0.082)\\
 MS1621+26                      & 0.427 & 14.977 (0.069)\\
 
% Photometric errors need to be recomputed
\hline
\end{tabular}
\end{table}

%\subsection{Combined sample}

The BCGs from the four samples (SpARCS, CNOC1, \citet{Stott2008}, and
\citet{Stott2010}) are combined into a single sample that is then used
to make three subsamples covering three broad redshift ranges: a
low-redshift subsample ($0.0 < z \le 0.3$), an intermediate-redshift
subsample ($0.3 < z \le 0.8$) and a high redshift subsample ($0.8 < z
< 1.65$).  The number of BCGs in each of these subsamples is listed in
Table~\ref{tab:results1}. We use these subsamples throughout the rest
of this paper.

\section{Estimating the stellar mass}

Following the methods used in previous works on determining the
stellar mass of BCGs \citep{Stott2008,Collins2009,Stott2010}, we use
the offset between the observed and predicted observer-frame Ks-band
magnitudes to estimate stellar mass. The predicted magnitude is
estimated from stellar population models that match the observer-frame
J-Ks colour of the BCGs over the entire redshift range covered by our
subsamples, i.e. from $z\sim 0$ to $z\sim1.6$. When converting between
luminosity and stellar mass we assume that the mass-to-light ratios of
the BCGs are independent of stellar mass.

\subsection{Modelling the J-Ks colour}\label{sec:colour}

We use the simple stellar population (SSP) models from
\citet[hereafter BC03]{Bruzual2003} to model the evolution of the spectral energy
distributions (SED) with cosmic time. There are a number of
ingredients that go into the models, such as the initial mass function
(IMF), the age and duration of the star burst, the metallicity of the
stars, and dust extinction. 

We assume a Chabrier IMF \citep{Chabrier2003}. The difference in the
resulting J-Ks colour from using a different IMF (for example, a
Salpeter IMF \citep{Salpeter1955}) is less than 0.02 mag over the
entire redshift range covered by the observations.  Similarly, the
stellar mass ratio between BCGs at low and high-redshift is relatively
unaffected by our choice of the IMF. The stellar masses themselves, however,
change significantly. Excluding stellar remnants, the difference is
about a factor of two for a given Ks-band luminosity. In this paper,
we do not use the stellar masses directly, just their ratios.

\begin{figure*}
 \includegraphics[width=16cm]{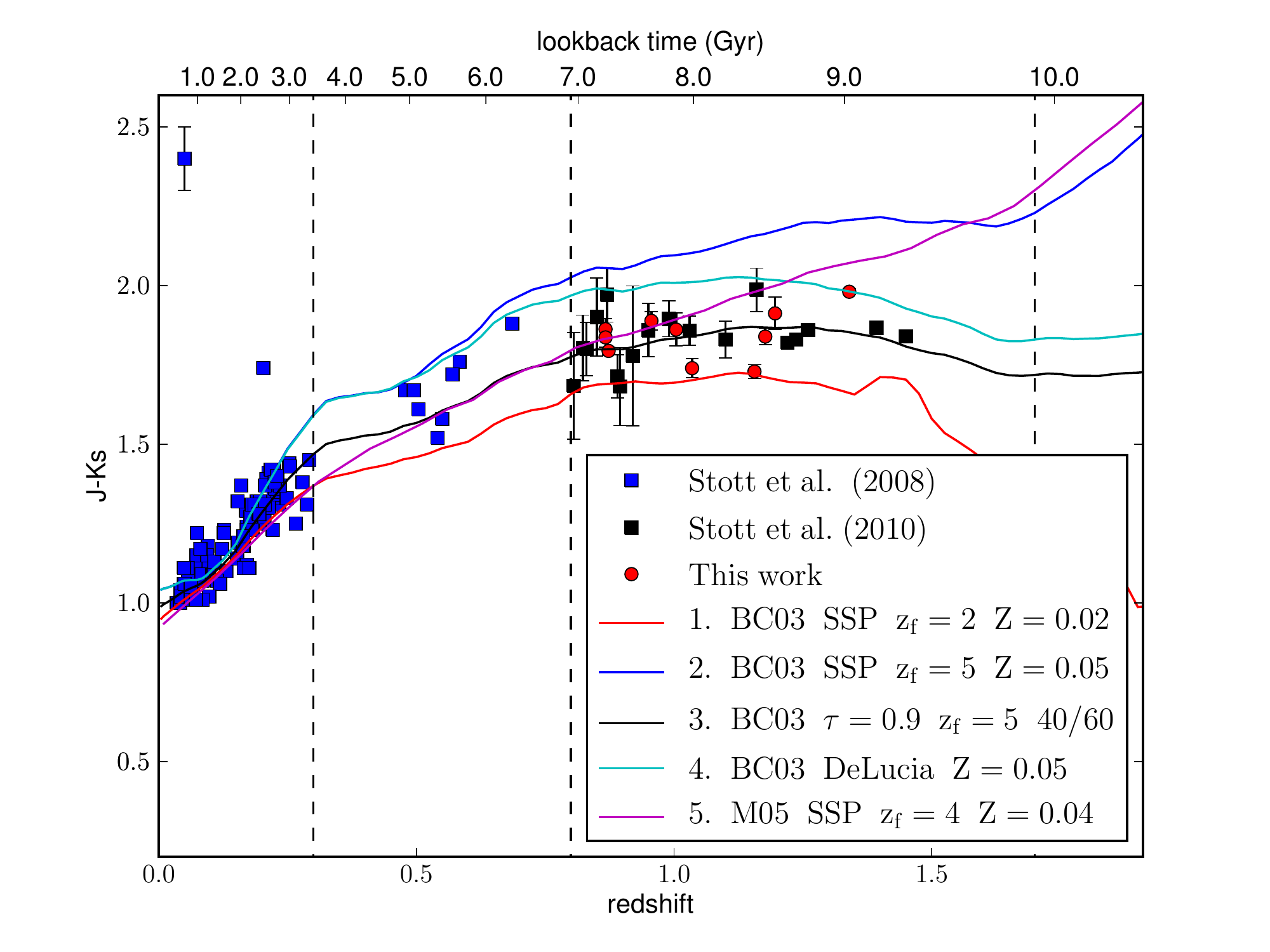} % Add redenning vectors
 \caption{The observer-frame J-Ks colour of BCGs in our sample as a
   function of redshift. The data from this paper are plotted as red
   circles. The two most distant clusters and clusters from the CNOC1
   sample are not plotted as they lack J-band data, The vertical
   dashed lines mark the boundaries of the low, intermediate and
   high-redshift subsamples that are described in the text. The
   evolution in the J-Ks colour for several stellar population models
   are plotted as the continuous lines. A broad range of models is
   shown. Note how well the our best fit model, model 3, which is the
   model we use to estimate stellar masses, describes the change in
   J-Ks colour with redshift. See text and Table~\ref{tab:models} for
   additional details.}\label{fig:models}
\end{figure*}

We assume that extinction from dust in negligible. From the small
amount of scatter in the colour of galaxies on the red-sequence, one
can infer that dust either reddens all galaxies by a small amount or
reddens a small number of galaxies considerably \citep{Meyers2012}. If
the former is true, then the amount of reddening affecting the BCGs in
our sample is unimportant. If the latter is true, then we would expect
to see significant colour outliers in Fig.~\ref{fig:models}, which we
do not see. 
% Of
%course, the redenning might be so great that one selects another
%galaxy to be the BCG.

%One can think of a situation in which the most massive galaxy has
%enough dust to make less bright than the second most massive galaxy,
%which would result in us selecting this galaxy as the BCG. Most of 
%the BCGs we select are centered in the center of the cluster. This
%seems a contrived situation, as it means that the 2nd most maaive
%galaxy needs to be cetered in the cluster too.

% Including dust tends to make the correction for dust larger at higher 
% redshifts, because the rest frame band passes shift to shorter wavelengths.
% This is just a guess. One would have to model it to see if this really is the
% case.

With the IMF set and dust ignored, we consider a series of models in
which we allow the star formation history to vary. We add an extra
dimension to these models by combining two models with identical
star-formation histories but different metallicities: a solar
metallicity model and a model that is two--and--a--half times
solar.  At low redshifts ($z \sim 0.03$), BCGs have
  metallicities that are around twice solar \citep{Loubser2009}.  We
add this extra degree of freedom because it is not possible to match
the colours of the BCGs over the entire redshift range
-- even by varying the star-formation history -- with the range of
metallicities available in BC03. We allow the mass ratio of the two
components to vary over the full range (i.e.~0.0 to 1.0) in steps of
0.1. We also allow the e-folding time, $\tau$, of the star-formation
rate to vary between 0.3 and 1.0\,Gyr in steps of 0.1\,Gy. We then
choose the model that best fits the data by finding the model with the
smallest chi-square. With the exception of the CNOC1 sample and the
two most distant clusters in the SpARCS sample, which lack J band
data, we use the entire sample when fitting the models. If an error in
the J-Ks colour is unavailable, we assume and error of 0.1 mag.

The best fitting model, model 3 in Fig.~\ref{fig:models}, has an
e-folding time of $\tau=0.9$\,Gyr and a composition that is split
60/40 between solar metallicity and a metallicity that is
two--and--a--half times higher. This model accurately describes the
general change in J-Ks colour with redshift although it does not
capture the scatter. In the remainder of this paper we use this model
to estimate stellar masses.

In order to demonstrate the sensitivity of the J-Ks colour on
metallicity and different star formation histories, we plot a series
of models. The parameters defining the models are listed
Table~\ref{tab:models}.  In model 1, all the stars form in a single
burst at $z=2$. In this model, all the stars have solar
metallicity. Model 2 is similar to model 1, except that we move the
burst to $z=5$ and increase the metallicity to two--and--a--half times
solar. In model 4, we move away from a single burst, using instead an
exponentially decaying burst of star formation with $\tau=1.03$\,Gyr
that starts forming stars at $z=10.6$. This model mimics the star formation history
of BCGs in the hierarchical models of \citet{DeLucia2007}, in which
50\% of the stellar mass is formed by $z=5$ and 80\% by $z=3$.  
%In later sections will use these models to explore
%how sensitive our conclusion on the history of mass assembly is to
%model selection. 
Our fifth model is a model from \citet{Maraston2005}. We will discuss
this model further in Section~\ref{sec:discussion}, where we will use
this model to test how sensitive out results are to our choice of
stellar evolution models.

\begin{table*}
\caption{Model parameters}\label{tab:models}
\centering           
\begin{tabular}{llllrrl}
\hline
Model &Description & Origin & IMF & Formation redshift & $\tau$ & Metallicity \\
&  & & & [Gyr] & \\
\hline
1&Low-redshift burst & BC03$^{a}$ & Chabrier & 2.0 & 0.0  & 0.02 (solar) \\
2&High-redshift burst&BC03 & Chabrier & 5.0 & 0.0  & 0.05 \\
3&Best fit model&BC03 & Chabrier & 5.0 & 0.9 & 60/40 split between 0.02 and 0.05 \\
4&DeLucia et al (2007) model &BC03 & Chabrier & 10.6 & 1.34 & 0.02 \\
5&High-redshift burst &M05$^{b}$ & Salpeter & 4.0 & 0.0 & 0.04\\
\hline
\multicolumn{5}{l}{Note a: \citet{Bruzual2003}}\\
\multicolumn{5}{l}{Note b: \citet{Maraston2005}}\\

\end{tabular}
\end{table*}

\subsection{Evolution in the stellar mass of BCGs}\label{sec:colour}

\begin{figure*}[ht]
\includegraphics[width=16cm]{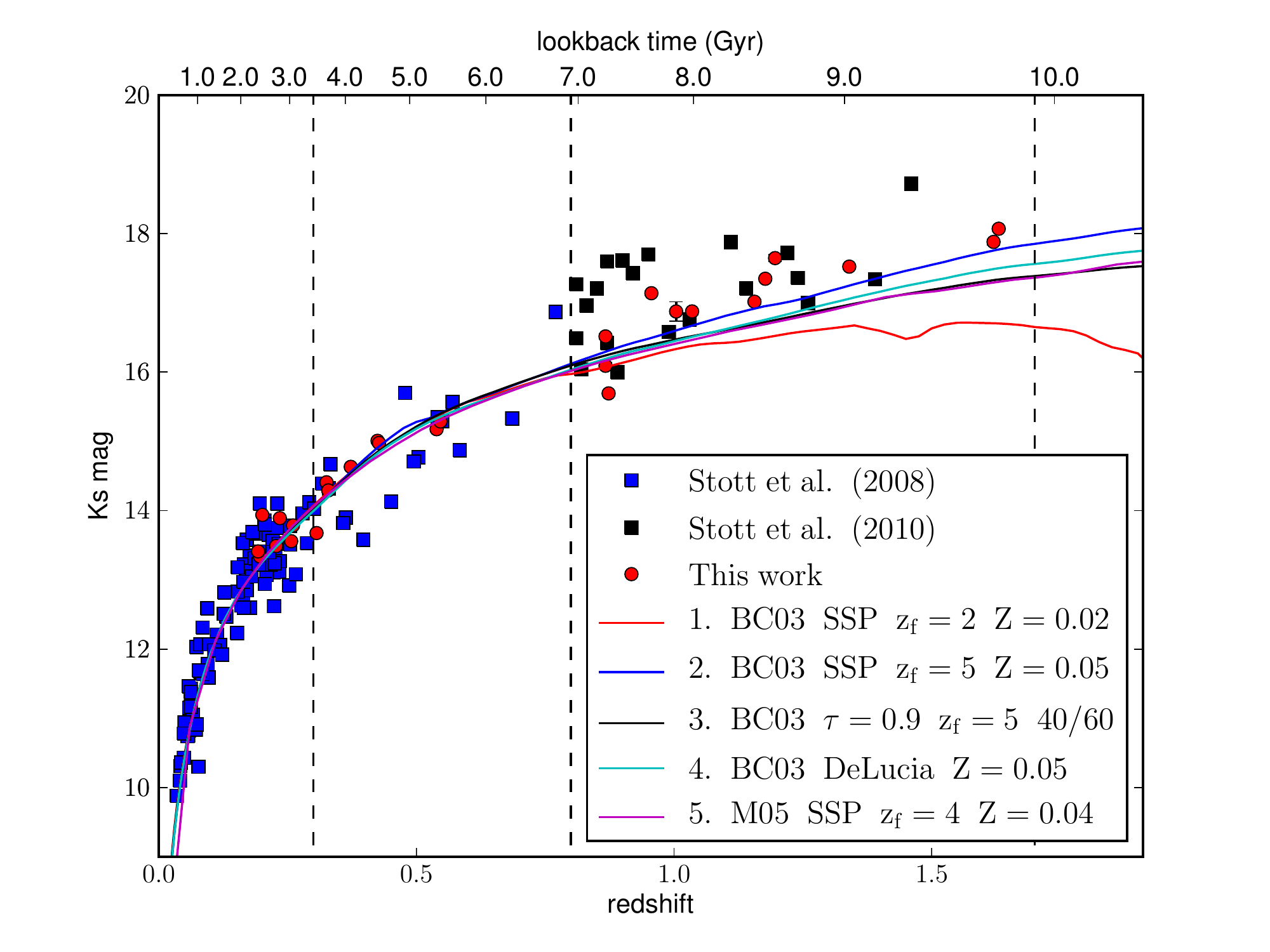}
\caption{The observer-frame Ks-band magnitude of BCGs as a function of
  redshift. The data from this paper are plotted as the red
  circles. Red circles beyond $z\sim 0.8$ are BCGs in the SpARCS
  clusters, while those below $z\sim 0.8$ are BCGs in the CNOC1
  clusters. The vertical dashed lines mark the boundaries of the low,
  intermediate and high-redshift subsamples that are described in the
  text. The predicted Ks magnitudes of the models plotted in
  Fig.~\ref{fig:models} and listed in Table~\ref{tab:models} are
  shown as the continuous lines. The models are normalised to the data
  in the low-redshift bin. They are discussed in
  Sec.~\ref{sec:colour}.  Note how all models tend to underpredict the
  flux in high-redshift BCGs. }\label{fig:Hubble}
\end{figure*}

In Fig.~\ref{fig:Hubble}, we plot the observer-frame Ks-band
magnitudes of the BCGs in our sample against their redshifts. BCGs from \citet{Stott2008} and
\citet{Stott2010} are plotted as the blue and black squares,
respectively, while BCGs in the SpARCS and CNOC1 clusters are plotted
as red circles. While the red circles generally land within the area covered by the
squares, the CNOC1 and SpARCS BCGs are less dispersed with respect to
the models than the BCGs in \citet{Stott2008}. 

Furthermore, the BCGs in the SpARCS clusters appear to be
  slightly brighter than the other BCGs in the high-redshift
  subsample, although this difference seems to be largely driven by a
  few BCGs at $z\sim 0.9$. The clusters hosting the SpARCS BCGs and
  the other clusters in the high redshifts subsample have similar
  masses, so the correlation between cluster mass and the BCG stellar
  mass (see Sec.~\ref{sec:masses}) is not the cause for the
  difference. Clusters in SpARCS were selected as galaxy
  overdensities, whereas most of the other clusters were selected
  through their X-ray emission. It is tempting to speculate that the
  difference is caused by the way the clusters were selected. However,
  the high-redshift sample is small, and we believe that a larger
  independent sample is required before one could conclusively state
  that sample selection is the reason for the difference.
 
In addition to the individual BCGs, we also plot the predictions of
the models described in the previous section. The
normalisation\footnote{ The normalisation is computed by matching the
  magnitude of the models at the median redshift of the low-redshift
  subsample with the median magnitude of the low-redshift
  subsample. It differs from the normalisation adopted in
  \citet{Collins2009} and \citet{Stott2010}. In these studies the
  normalisation occurs over a more restrictive redshift interval
  ($z < 0.05$).  See Sec.~\ref{sec:comp} for further details.}  of the
models is constrained by the data in the low-redshift subsample. With
this normalisation, the BCGs at low redshift correspond to galaxies
that are about 2\,mag brighter than a $L_{\star}$ galaxy in the Coma
cluster \citep{dePropris1998}.

The stellar mass of individual galaxies is derived by converting the offset in
magnitude between model 3, the model that best describes the evolution
in the J-Ks colour with redshift, and the observed Ks-band magnitude
to a stellar mass. We normalise the stellar mass of the BCGs to the
stellar mass they would
have by today, using the modelled decrease in stellar mass
with time from stellar winds and supernova explosions
\citep{Bruzual2003}. The results for the model that best follows the
general evolution in the J-Ks colour, model 3, are shown in
Table~\ref{tab:results1}. In the last column of this table, we also
list the median stellar mass of the BCGs at the cluster redshift. When
comparing the stellar mass of BCGs at low and high redshift we compare the
stellar masses they would have by today - the second last column in
this table.

\begin{table*}
\caption{The three subsamples described in Sec.~\ref{sec:samples} and
  the SpARCS sample. For all quantities, we report the median value.}\label{tab:results1}
\centering           
\begin{tabular}{lrlccccc}
\hline
Subsample & Redshift range         & Size & Redshift &
Cluster mass$^a$ & Cluster Mass$^b$ & BCG mass$^c$ &  BCG mass$^d$\\        
            &                                &                    &
            &  [$10^{15}\,M_{\odot}$]    &  [$10^{15}\,M_{\odot}$]  & [$10^{12}\,M_{\odot}$] &
             [$10^{12}\,M_{\odot}$]\\
\hline
Low                    & $ z \le 0.3$           & 93 (90) & 0.17 & 0.79$^e$ & 0.59$^e$& 0.45 & 0.46\\
Intermediate      & $ 0.3 < z \le 0.8$  & 25 (18) & 0.45 & 2.34$^e$ & 1.27$^e$ & 0.50 & 0.52\\
High                  & $ 0.8 < z \le 1.7$  & 32 (32) & 1.00 & 1.20 & 0.30 & 0.29 & 0.32\\
SpARCS             &$ 0.8 < z \le 1.7$    & 12 (12) & 1.10 & 1.19 & 0.29 & 0.31 & 0.34\\
\hline
\multicolumn{7}{l}{Note a: Cluster masses corrected for
 the growth they are likely to have by today.}\\
\multicolumn{7}{l}{Note b: Cluster mass at the redshift of the cluster.}\\
\multicolumn{7}{l}{Note c: The stellar mass of the BCG at redshift zero (accounts for stellar mass loss).}\\
\multicolumn{7}{l}{Note d: The stellar mass of the BCG at the redshift of the cluster.}\\
\multicolumn{7}{l}{Note e: Computed for the subset of clusters (numbered
  in brackets) with masses. See text for details.}
\end{tabular}
\end{table*}

Without making any correction for the positive correlation between the
stellar mass of the BCG and the mass of cluster \citep[see][and the
next section]{Edge1991,Burke2000,Brough2008,Whiley2008}, the data
indicate that the stellar mass of the BCGs increase by a factor of
\growthunmatched\
between $z\sim1$ and $z \sim 0.17$.  The errors are determined by
bootstrap resampling. The increase is found for the
high-redshift subsample as a whole and for a smaller subsample
consisting of just the clusters from SpARCS.  In the next section we
examine how the correlation between the stellar mass of the BCG and
the mass of cluster affect our results.

\subsection{Accounting for cluster masses}\label{sec:masses}

The stellar mass of BCGs correlates with cluster mass in the sense
that larger clusters tend to host larger BCGs
\citep{Edge1991,Burke2000,Brough2008, Whiley2008, Stott2012}. Comparing
the stellar mass of BCGs in our three subsamples
without accounting for this correlation will lead to biased results if
the median mass of the clusters in the subsamples differ.

To account for this correlation, we first need to estimate how
clusters grow in mass so that we can fairly compare clusters that are
observed at different redshifts. Over the redshift range that our
sample covers, clusters grow significantly. For example, in the
hierarchical model of structure formation, a cluster with a mass of
$5\times10^{14}\,M_{\odot}$ at $z=1$ is predicted to grow by a factor
of about three by today
  \citep{Wechsler2002,Fakhouri2010}.  \citet{Fakhouri2010}, who use
  the Millennium and Millennium II simulations
  \citep{Springel2005,Boylan2009} and \citet{Wechsler2002}, who use an
  independent simulation \citep{Bullock2001}, find similar growth
  rates. We use mean accretion rates in \citet{Fakhouri2010} to
estimate the mass each cluster should have by the current epoch using
the masses they had at the redshifts they were observed. We describe
how we estimate cluster masses at the redshift at which they were
observed in Sec.~\ref{sec:ClusterMass}.

After evolving our clusters forward in time to today, we find that the
median mass of the clusters in our three subsamples differ by as much
as a factor of three (see Table \ref{tab:results1}). The differences
in the subsamples reflect the volumes probed and sensitivity limits of the
surveys that were used to build our subsamples.  Since clusters in the
intermediate and high-redshift subsamples are, by the current epoch,
more massive than those in the low-redshift subsample, the correlation
between cluster mass and BCG stellar mass -- if uncorrected -- leads
to an underestimate in the amount of evolution in the stellar mass of
BCGs.

In this paper, we explore a couple of approaches to account for the
correlation. In the first approach, we first match the cluster mass
distributions in the samples being compared before comparing the
masses of the BCGs. In the second approach, we normalize BCG stellar
masses to some fiducial mass using the relationship between cluster
mass and the BCG stellar mass. 

Our approaches to account for this correlation differs from approaches
used in the past. In \citet{Whiley2008}, clusters are grouped
according to the mass they had at the redshift they were observed. In
\citet{Stott2010}, the mass of BCGs are compared to the mass of BCGs
from semi-analytic models after first matching the masses of the
clusters in the semi-analytic models to the observed masses.

\subsubsection{Estimating Cluster mass }\label{sec:ClusterMass}

Given the heterogeneous nature of the data that is available for our
clusters, we estimate cluster masses\footnote{We use $M_{200}$ for
  cluster masses. $M_{200}$ is the mass contained within a radius
  within which the mean density of the cluster exceeds the critical
  density of the Universe at the redshift of the cluster by a factor
  of 200.} in
different ways. For clusters in the low and intermediate-redshift
subsamples of \citet{Stott2008}, we use the $M_{500}$ masses listed in
\citet{Mantz2010}. These masses are converted to $M_{200}$ assuming
that the cluster mass profile follows a Navarro-Frenk-White profile
\citep{Navarro1997} with a concentration index of 5. For the
conversion, we use the formulae listed in Appendix C of
\citet{Hu2003}. Only about a quarter of the clusters in
\citet{Stott2008} are in \citet{Mantz2010}. To increase the number of
clusters in \citet{Stott2008} with mass measurements, we use the X-ray
temperatures \citep{Ebeling2007,Ebeling2010} and the X-ray
luminosities \citep{Ebeling1996,Ebeling1998,Ebeling2000} of these
clusters (the luminosities were corrected for the cosmology used in
this paper) and the temperature--mass and luminosity--mass relations
in \citep{Mantz2010} to estimate cluster masses.
%In total, 93 of the
%low-to-intermediate redshift clusters that are listed in
%\citep{Ebeling1996,Ebeling1998,Ebeling2000,Ebeling2007,Ebeling2010,Mantz2010}
%are used.

Twenty one of the clusters in
\citep{Ebeling2007,Ebeling2010,Mantz2010} are also in
\citep{Ebeling1996,Ebeling1998,Ebeling2000}, which enables us to
compare the mass derived from the X-ray luminosity with the mass
derived from X-ray temperature. The median ratio is 1.14 with a
s.d. of 0.53. We correct the masses determined from the X-ray
luminosity by the median ratio and use the s.d. as a measure of the
uncertainty in the conversion.

For the SpARCS clusters, we estimate the cluster mass from the
line-of-sight velocity dispersion (Wilson et al., in preparation). For
clusters in the high-redshift subsample of \citet{Stott2010}, we use
the X-ray temperature reported in that paper and convert them to
masses using the relation in \citet{Mantz2010}.  For all but two of
the clusters in the intermediate-redshift subsample of
\citet{Muzzin2007a} we use the X-ray temperatures listed in
\citet{Hicks2006} and convert them to masses using the relation in
\citet{Mantz2010}. For the remaining two clusters (MS1224+20 and
MS1231+15), we use the masses listed in \citet{Muzzin2007b}, which are
computed from the line-of-sight velocity dispersion.

After removing clusters from \citet{Stott2008} that are not listed in
\citet{Ebeling1996,Ebeling1998,Ebeling2000,Ebeling2007,Ebeling2010}
and \citet{Mantz2010}, we end up with 90 and 18 clusters in the low-
and intermediate-redshift subsamples, respectively.  The number of
clusters in the high-redshift subsample is unchanged. The numbers are
listed in Table~\ref{tab:results1}. Altogether, there are 152 clusters
in our three subsamples. 

\subsubsection{Cluster mass vs. BCG stellar mass}\label{sec:clusterVsBCG}

\begin{figure}
\includegraphics[width=8cm]{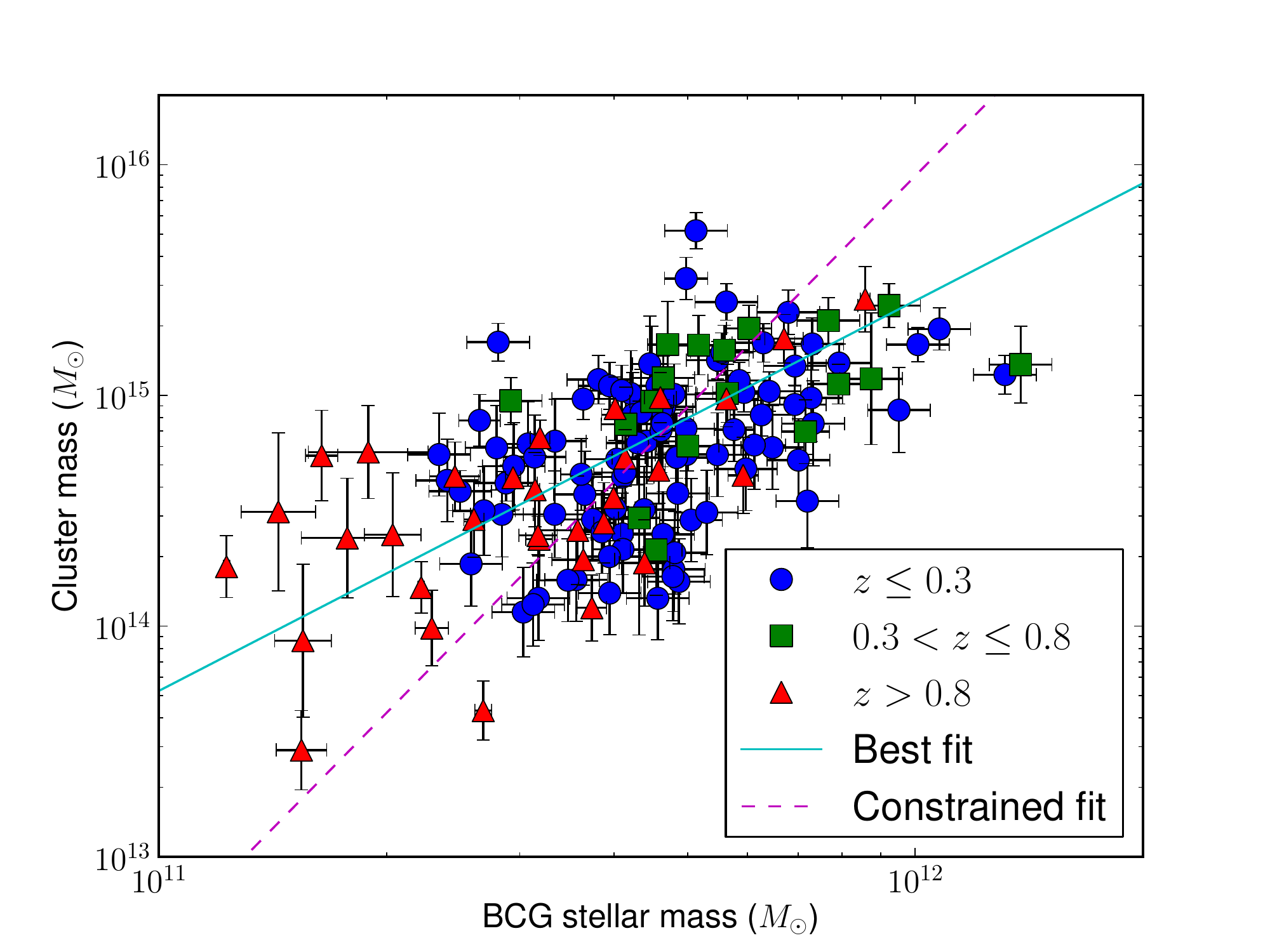}
\caption{The correlation between the mass of the cluster at the epoch
  at which it was observed and the stellar mass of the BCG. The
  different symbols represent different redshift ranges. The solid
  line is a fit to the data. Setting the index of the power law to the
  value reported in \citet{Hansen2009} results in a poorer fit to the
  data (dashed line).}\label{fig:correlation}
\end{figure}

The correlation between cluster mass (at the redshift at which it was
observed) and BCG stellar mass for these three subsamples is shown in
Fig.~\ref{fig:correlation}. Errors in the mass of the BCGs are derived
from errors in the photometry. If an error in the Ks-band photometry
was unavailable, we conservatively set the error to 10\%. Our results
are not very sensitive to this value, as errors in the cluster masses are
much larger. Errors in the mass of the clusters are discussed in
\ref{sec:ClusterMasses}. We note that clusters in the
intermediate-redshift subsample generally have higher masses than
clusters in the low-redshift subsample. As noted earlier, the difference between
subsamples reflect the volumes probed and the sensitivity limits of
the individual surveys that were used to build the subsamples.

We fit a power law to the data, using a lognormal distribution to
represent the likelihood of getting a certain data point given the
model and allowing for additional dispersion by scaling the
measurement uncertainties.  The index of the power law that
corresponds to the maximum of the posterior distribution is \fitindex.
Because we treat errors and the amount of extra dispersion in both
axes equally, our results are robust to flipping the axes in the
fit. The amount of extra dispersion found in the fit corresponds to
increasing the size of the error bars by a factor of 1.5.

The index of the power law suggests that clusters accrete mass five
times faster the BCGs accrete stellar mass. Within uncertainties, the
index is similar to that found in \citet{Stott2010}, who find $2.4 \pm
0.6$ and \citet{Stott2012} who find $1.3 \pm 0.1$. Some of the
difference between our results and those in
\citet{Stott2010,Stott2012} come from the way the samples are selected
and the way the analysis is performed.  Our best fit index is about a factor
of two smaller than those reported in earlier works \citep{Lin2004,
  Popesso2007, Brough2008, Hansen2009}. For example,
\citet{Hansen2009}, find an index of $3.3$ between the i-band
luminosity (k-corrected to $z=0.25$) and $M_{200}$. We redid the fit
with the index constrained to this value.  The resulting relation is
shown in Fig.~\ref{fig:correlation} as the dashed line. It is a poorer
fit to the data.

Our fit to the entire sample seems to be largely driven by the
clusters in the high-redshift subsample, whereas most of the clusters
in \citet{Hansen2009} were at low redshift. This raises the
possibility that there is evolution in the index of the power law with
redshift. Alternatively, the difference might be caused by
redshift-dependent selection effects. In accounting for cluster masses
in the following sections, we adopt a conservative approach and
examine how our results depend on which index we choose to use. We
will find that our conclusions are robust to this choice.

\subsubsection{Accounting for cluster masses }\label{sec:ClusterMasses}

As foreshadowed earlier, we use two approaches to account for the
correlation between cluster mass and BCG stellar mass. We discuss the
first approach in this section and discuss the second approach in the
section that follows.

In the first approach, we randomly select clusters from the three
subsamples until the mass histograms\footnote{We use a bin size of $2
  \times 10^{14}$ and we use the mass the clusters are likely to have
  by today.} of the subsamples match. Clusters are matched according
to the mass they will have by the current epoch.  Implicit in this
approach is the method we use to estimate how clusters build up their
mass with time.

\begin{figure}
\includegraphics[width=8cm]{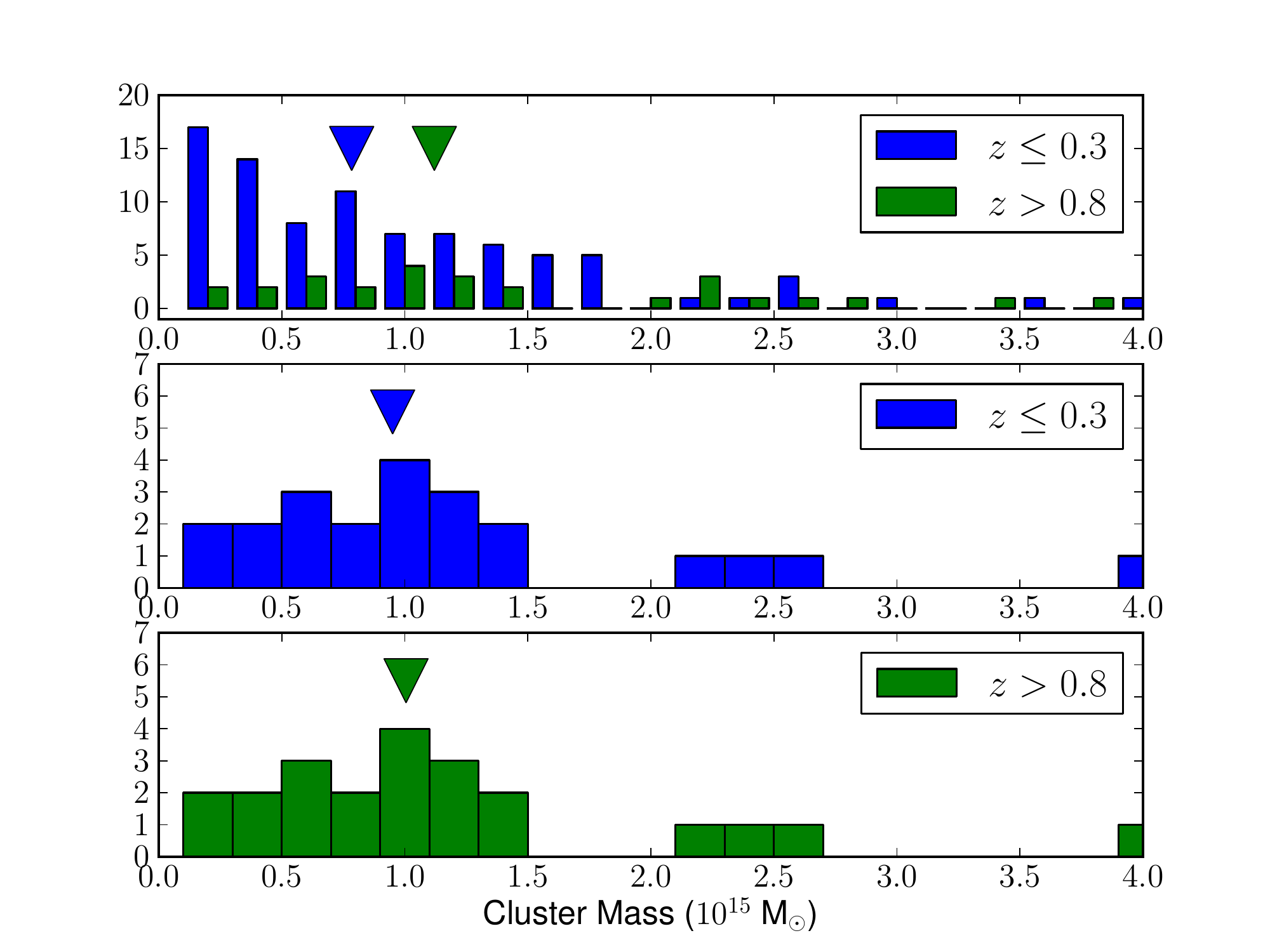}
\caption{{\bf Upper Panel} A histogram showing the distributions of
  cluster mass (extrapolated to the current epoch) for the low and
  high-redshift subsamples. The median masses are marked with the
  downward pointing arrows. Note how the median mass of the two
  distributions differ and how skewed the low redshift subsample is
  with respect to the high-redshift one.  {\bf Lower two panels}
  Histograms of the re-sampled low and high-redshift subsamples. They
  are resampled so that they are identical for a bin width of
  $2 \times 10^{14}$\,M$_{\odot}$. The median masses, marked with the downward
  pointing arrows, are now more similar. There are 23 objects in the
  lower two histograms.}\label{fig:masshist}
\end{figure}

We cannot match all three subsamples simultaneously, because trying to
get all the histograms to match would result in very few objects per
subsample. Instead we compare the low-redshift subsample with the
intermediate and high-redshift subsamples separately. The method is
illustrated in Fig.~\ref{fig:masshist} for the comparison between the
low and high-redshift subsamples.

%We investigated applying a lower and a upper mass cut to the subsamples
%without trying to balance the distributions. However, the low redshift
%subsample is too heavily skewed to low cluster masses. Conversely, the
%high-redshift subsample was too heavily skewed to high cluster masses. It
%was not possible to match the median mass of the clusters in the three
%subsamples with this approach.

In order to get a measure of the uncertainties in the derived mass
ratios, we do two things. We first perturb the cluster mass by an amount that depends on two sources of
error: the uncertainty in the measurement of the mass proxy (X--ray
temperature, X--ray gas mass, X-ray luminosity or line--of--sight
velocity dispersion) and the intrinsic scatter between the mass proxy
and the mass.  For masses that are determined from the X-ray gas mass
or the X-ray temperature, we assign a scatter of 15\%
\citep{Mantz2010}. For masses determined from the line-of-sight
velocity dispersion, we assign a scatter of 30\% \citep{Hicks2006}.
For masses inferred from the X--ray luminosity, we use 50\%, which we
derived earlier. The magnitude of the perturbation is drawn from a
lognormal distribution. The s.d. of the distribution is set equal to
the two uncertainties added in quadrature.

\begin{figure}
\includegraphics[width=8cm]{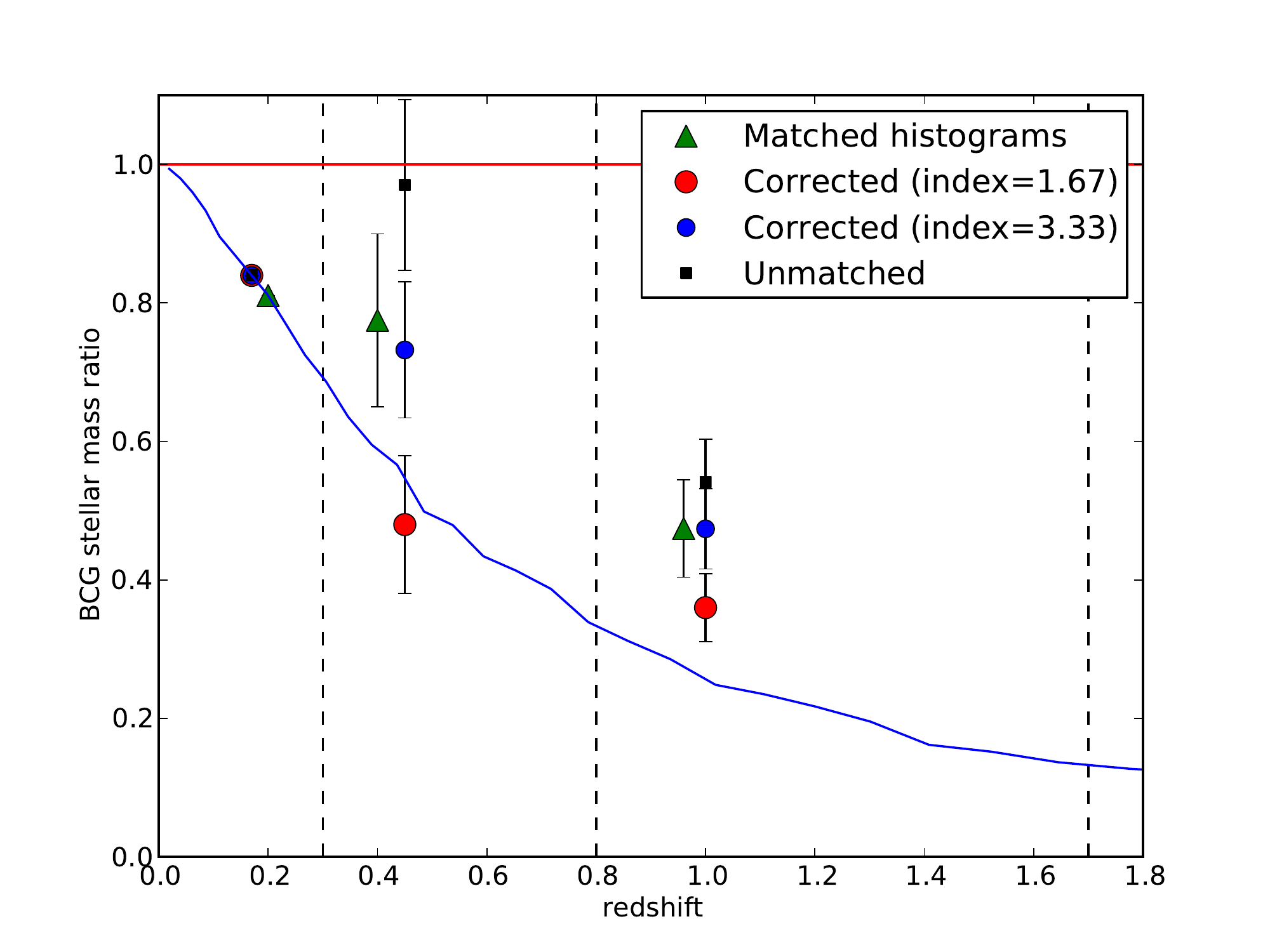}
\caption{The evolution in the median stellar mass of BCGs as a
  function of redshift. The green triangles take into account the
  correlation between cluster mass and the stellar mass of its BCG by
  matching clusters according to the masses they will have by the
  present epoch. In a second approach, the red and blue circles
  account for this correlation using the relations shown in
  Fig.~\ref{fig:correlation}. The small black squares do not account
  for this correlation. Note how all the points in the high-redshift
  bin lie below the red line, how the green, blue and red points in
  the intermediate and high-redshift bins lie below the black points,
  and how these points are a better match to the \citet{DeLucia2007}
  model (solid blue line).  All points are normalised so that their
  low-redshift points land on this model. The vertical dashed lines
  mark the boundaries of the low, intermediate and high-redshift
  subsamples that are described in the text. The points are plotted at
  the median redshifts of the subsamples. They differ slightly between
  the green, red, blue, and black points because a more restricted
  range of clusters is selected when matching cluster masses. See text
  for details on how the error bars are computed. The red horizontal
  line represents no mass evolution. The data used in this plot are
  summarised in Table~\ref{tab:summary}}\label{fig:massevol}
\end{figure}

Secondly, we resample the three subsamples with replacement (bootstrap
resampling) to allow for uncertainties that come from sample
size. Only then do we try to match the histograms in the three
subsamples. We repeat this exercise 100 times for each comparison to
create 200 realisations from the data. For each realisation, we
compute the median BCG stellar mass, the median cluster mass and
median redshift. For each comparison, we then average the results from
the 100 realisations and get an estimate of the robustness of the
results from the variance. The results of the comparisons are listed
in Table~\ref{tab:comparison} and shown in Fig.~\ref{fig:massevol}.
The uncertainty in the last column in Table~\ref{tab:comparison} is
computed from the 100 realisations and gives an indication of the
robustness of the result. The uncertainties are plotted as the
vertical error bars in Fig.~\ref{fig:massevol}. No other uncertainties
are included in these error bars.

Between $z\sim 0.9$ and $z \sim 0.2$, the stellar mass of BCGs increase by a
factor of \growth. This is larger than the increase
reported in the previous section, which did not account for the
correlation between cluster mass and the stellar mass of the BCG.  

\begin{table*}
  \caption{BCG mass ratios. The cluster mass distributions are matched using the masses the clusters
    will have by the current epoch.}\label{tab:comparison}
\centering           
\begin{tabular}{l|ll|ll|ll|l}
\hline
Samples  a and b        & \multicolumn{2}{c}{Median Redshifts}
&\multicolumn{2}{c}{Median cluster masses} & \multicolumn{2}{c}{Median
 BCG  masses}  & BCG mass ratio$^1$\\        
                           & Sample a & Sample b & Sample a & Sample
                           b & Sample a & Sample b \\
                           &                          &&
                           \multicolumn{2}{c}{[$10^{15}\,M_{\odot}$]}
                           &
                           \multicolumn{2}{c}{[$10^{12}\,M_{\odot}$]}
                           & \\
\hline
   low (a) -- intermediate (b) &  0.20 & 0.40 & 1.59 & 1.61 & 0.54 & 0.51 & 0.96 $\pm$ 0.20 \\
   low (a) -- high (b) &  0.17 & 0.97 & 0.84 & 0.84 & 0.45 & 0.26 & 0.58 $\pm$ 0.08 \\

\hline
\multicolumn{8}{l}{Note 1: Defined as the median
 stellar mass of subsample b divided by the median stellar mass of subsample a} \\
\end{tabular}
\end{table*}

We repeated our analysis by comparing the stellar mass of BCGs in
clusters that have the same mass at the redshift they were
observed. This is the method used in \citet{Whiley2008}. The results
are presented in Table~\ref{tab:comparison2}. Not surprisingly, due to
the correlation between the stellar mass of the BCG and the mass of
the cluster, and the considerable growth in cluster mass between
$z\sim1$ and today, the evolution in the stellar mass of the BCG is less
evident when clusters are compared in this way.

\begin{table*}
  \caption{As for Table~\ref{tab:comparison} with the difference that
    the matching is done using the masses the clusters have at the
    redshift at which they were observed.}\label{tab:comparison2}
\centering           
\begin{tabular}{l|ll|ll|ll|l}
\hline
Samples a and b    & \multicolumn{2}{c}{Median Redshifts}
&\multicolumn{2}{c}{Median cluster masses} & \multicolumn{2}{c}{Median
 BCG  masses}  & BCG mass ratio$^1$\\        
                           & Sample a & Sample b & Sample a & Sample
                           b & Sample a & Sample b \\
                           &                          &&
                           \multicolumn{2}{c}{[$10^{15}\,M_{\odot}$]}
                           &
                           \multicolumn{2}{c}{[$10^{12}\,M_{\odot}$]}
                           & \\
\hline
   low (a) -- intermediate (b) &  0.20 & 0.44 & 1.02 & 1.06 & 0.51 & 0.51 & 1.02 $\pm$ 0.16 \\
   low (a) -- high (b) &  0.10 & 1.00 & 0.33 & 0.30 & 0.41 & 0.29 & 0.71 $\pm$ 0.10 \\

 \hline
\multicolumn{8}{l}{Note 1: Defined as the median
 stellar mass of subsample b divided by the median stellar mass of subsample a} \\ 
\end{tabular}
\end{table*}

\begin{table}
\caption{A summary of the data appearing in Fig.~\ref{fig:massevol}}\label{tab:summary}
\centering           
\begin{tabular}{lll}
\hline
Method & Redshift & Mass Ratio$^a$ \\
\hline
Matched histograms &0.20 & 0.81\\
 & 0.40 & 0.77 $\pm$ 0.12\\
 & 0.96 & 0.47 $\pm$ 0.07\\
\hline
No matching &0.17 & 0.84\\
 & 0.45 & 0.97 $\pm$ 0.12\\
 & 1.00 & 0.54 $\pm$ 0.06\\
\hline
Correcting with index 1.67 &0.17 & 0.84\\
 & 0.45 & 0.48 $\pm$ 0.10\\
 & 1.00 & 0.36 $\pm$ 0.06\\
\hline
Correcting with index 3.33 &0.17 & 0.84\\
 & 0.45 & 0.73 $\pm$ 0.10\\
 & 1.00 & 0.47 $\pm$ 0.06\\
\hline

\multicolumn{3}{l}{Note a: The ratios are scaled so that the
  low-redshift point}\\
\multicolumn{3}{l}{matches the prediction of \citet{DeLucia2007}.}\\
\end{tabular}
\end{table}

\subsubsection{An alternative approach}\label{sec:alternative}

An alternative approach to account for the correlation between cluster
mass and BCG stellar mass is to adjust the BCG stellar mass according
to the relation shown as the solid line in
Fig.~\ref{fig:correlation}. As in the first approach, we use the cluster mass
extrapolated to the current epoch and not the mass
they had at the epoch they were observed.  The results are shown as
the red points in Fig.~\ref{fig:massevol}. The errors are derived
using bootstrap resampling.

Compared to the previous approach, we find stronger growth in the BCG
stellar mass as a function of redshift and better agreement between
the data and the semi-analytic models of \citet{DeLucia2007}. However,
the result depends on the relation shown in
Fig.~\ref{fig:correlation}.  Adopting the relation found by
\citet{Hansen2009} instead of the relation we find, for example,
results in less growth (the blue points in
Fig.~\ref{fig:massevol}). The point at intermediate redshifts is
affected most, since these clusters will be, by today, three times
more massive than clusters in the low-redshift sample (see
Table~\ref{tab:results1}), thereby leading to significant adjustments. The
point at high-redshifts is affected less, because these clusters will
be, by the current epoch, similar in mass to clusters in the low-redshift sample.

We do not adopt the relation found in \citet{Hansen2009}. Instead we
use it to demonstrate the sensitivity of the approach to changes in
the power-law index. The sample used in \citet{Hansen2009} to compute
the relation is restricted to clusters in the redshift range $0.1 \le
z \le 0.3$. There are also differences in the analysis. Cluster masses
in \citet{Hansen2009} are estimated from the optical richness and BCG
masses are estimated from the observer--frame i--band, which, when
combined with the optical selection, may lead to biases that influence
the result.

%The alternative approach makes a couple of implicit assumptions.  In
%addition to assuming how clusters build up their mass with time, which
%is common to both approaches, we make the assumption that the stellar
%mass of all the BCGs, independent of their redshift or the mass of
%their host cluster, can be corrected with the same
%relation power-law relation. 

We use the differences in the results between this approach (using the
best fit power law index), the approach
described in Sec.~\ref{sec:ClusterMasses}, and the approach of not
applying any correction as an estimate of the size of the systematic
error. Clearly, between the intermediate and low-redshift samples, the
evidence for evolution is marginal. The statistical and systematic
uncertainties are too large.

However, between the high and low-redshift samples, the evidence for
evolution is clear and unambiguous. Between $z\sim 0.9$ and $z \sim
0.2$, the stellar mass of BCGs increase by a factor of \growth\ with a
spread of 0.4 spanning the three approaches.

%Nevertheless, irrespective of which approach we choose to correct for
%the correlation between cluster mass and the stellar mass of the BCG,
%including the option of applying no correction, we find significant
%evolution in the stellar mass of BCGs between $z\sim 0.9$ and $z \sim
%0.2$.

\section{Discussion}\label{sec:discussion}

The semi-analytic model of \citet{DeLucia2007} predicts that BCGs grow
by a factor of almost three in stellar mass between $z\sim0.9$ and
$z\sim0.2$. Over the same redshift interval, we observe that BCGs
increase their stellar mass by a factor of \growth.

Our result depends on the methods we have used to analyse the data
and the choice of models that we have used to estimate stellar masses. We
discuss each of these in turn, finding that our results are robust.

\subsection{Estimating the Ks band flux}

Throughout this paper we have estimated the Ks-band flux using {\tt
  MAG\_AUTO} in SExtractor, which is a Kron-like magnitude
\citep{Kron1980} within an elliptical aperture.  Undoubtably, {\tt
  MAG\_AUTO}, like all other measures of the total magnitude used in
the literature, will be systematically biased to low or high values
depending on the nature of the object being measured. The bias can
occur for a number of reasons, such as the number and brightness nearby
neighbours, the presence of intra-cluster light and/or a cD envelope,
the point spread function (i.e.~seeing) and residual errors that come
from the imprecise removal of the bright night sky from near-IR
images. For the purpose of comparing the stellar mass of BCGs at low
and high redshifts, the most important aspect of the measurement is
that the bias does not change with redshift. In this section we search
for evidence of the bias changing with redshift and try to estimate
how large this bias may be.

%examine how large this bias could be by using the integrated magnitude
%from {\tt
% GALFIT}\footnote{http://users.obs.carnegiescience.edu/peng/work/galfit/galfit.html}
%\citep{Peng2002} to estimate magnitude.

\citet{Graham2005} noted that Kron-like magnitudes can significantly
underestimate the flux of galaxies with S\'{e}rsic profiles.  We ran
our own investigation into the accuracy of {\tt MAG\_AUTO} by
inserting objects with S\'{e}rsic profiles in simulated images. In the
simulations, we mimicked the background noise and image quality of the
real data. For de Vaucouleur profiles\footnote{The de Vaucouleur profile is
  equivalent to a S\'{e}rsic profile with the S\'{e}rsic index, n, set
  to 4.}, {\tt MAG\_AUTO} misses between 18\% and 35\% of the
flux, depending on the redshift. The trend with redshift is non-monotonic. At $z=0.1$, 25\% of
the flux is missed. This decreases to 18\% by $z=0.4$, then increases
to 25\% by $z=1.0$ and to 35\% by $z=1.6$. For higher S\'{e}rsic
indices (we tested indices as high as $n=8$), higher fractions of the
flux are missed by {\tt MAG\_AUTO}; however the trend with redshift is
the same.

% Compare between redshift 0.2 and 0.9

Clearly, if the profiles of BCGs evolve with time, then
  there will a redshift-dependent bias in the stellar masses that are
  inferred from the photometry. For example, if low-redshift BCGs had
de Vaucouleur profiles and high-redshift BCGs had S\'{e}rsic profiles
with $n=8$, then we would {\it overestimate} the flux of the low
redshift BCGs relative to their distant cousins, and therefore their
stellar mass, by around 10\%. If the opposite was true
(i.e.~high-redshift BCGs had de Vaucouleur profiles and low-redshift
BCGs had S\'{e}rsic profiles with $n=8$), then we would {\it
  underestimate} the flux of the low redshift BCGs by 3\%. The
asymmetry is caused by the dependence of how accurately {\tt
  MAG\_AUTO} measures total magnitude with redshift and
  seeing.  Observational constraints on the redshift dependence of
  the S\'{e}rsic index show that the redshift dependence is much
  weaker than the range of values that we have considered here
  \citep{Stott2011}.

In our simulations, we neglected errors in the photometry that come
from nearby (in projection) galaxies and imprecise sky-subtraction.
To investigate these issues, we compare the integrated light profiles
of the BCGs measured with version 3.0.4 of
GALFIT\footnote{http://users.obs.carnegiescience.edu/peng/work/galfit/galfit.html}
\citep{Peng2002}. Galaxies neighbouring the BCG were either fitted
simultaneously (if they were within 2-3\arcsec\ of the BCG) or masked
as bad pixels (if they were further than this). We used the residual
images and the reduced $\chi^2$ of the fit to determine how well the
data were described with the model. With the exception of SpARCS-0035
and SpARCS-1638, the BCGs could be modelled satisfactorily (a reduced
$\chi^2$ close to one) with a de Vaucouleur profile. The BCGs of both
SpARCS-0035 and SpARCS-1638 were better fit with S\'{e}rsic profiles
that had a higher S\'{e}rsic indices.

For the SpARCS clusters, which have a median redshift of $z \sim 1.1$,
the offset between {\tt MAG\_AUTO} and the magnitude determined by
integrating the fitted GALFIT profile out to infinity has a median
value of 0.30 mag (i.e.~relative to the integrated GALFIT flux, {\tt
  MAG\_AUTO} underestimates the flux. For clusters in the CNOC1
sample, which have a median redshift of $z \sim 0.28$, the offset
between the integrated GALFIT magnitude and {\tt MAG\_AUTO} has a
median value of 0.49 mag, which is considerably larger than the median
value found for the SpARCS clusters.
%If we make the assumption that the integrated GALFIT
%magnitude is unbiased, then correcting for the bias strengthens the
%case for significant evolution.

%There are a number of reasons why the difference is smaller for the
%SpARCS BCGs.  
In part, the difference between the offsets is due to the way the
aperture in {\tt MAG\_AUTO} is defined. The size of the aperture
depends on the seeing convolved profile of the BCG. Since the CNOC1
and SpARCS samples were taken in similar seeing, the apertures for the
SpARCS BCGs are affected more by the seeing, since the angular size of
the BCGs relative to the seeing disk is smaller. This leads one to
using apertures for the SpARCS BCGs that are larger than one would
have used if the ratio of the seeing to the angular size of the BCG
was the same for both samples.  This then leads to a smaller
difference for SpARCS BCGs.

%The smaller difference, may also be due to the way the near-IR data have been
%processed. The diameters of the apertures used by {\tt MAG\_AUTO} are
%quite large: the physical diameter averages around 55\,kpc for both
%samples. A small error in the sky-subtraction may have a significant
%impact on the derived magnitudes and this would affect {\tt MAG\_AUTO}
%and the integragted GALFIT magnitude in different ways.

%Finally, the smaller difference may be due to evolution in BCG profiles,
%although we found earlier in this section that the effect is likely to
%be small or have the wrong sign.

The offset between the differences translates directly into a relative
offset in the stellar masses of the BCGs in SpARCS and CNOC1. Relative
to the stellar masses of the BCGs in the SpARCS clusters, we are
underestimating the stellar masses of the BCGs in the CNOC1 clusters
by a factor of 1.2. If this offset were applicable to the rest of the BCGs in our
low-redshift subsample, then we would be underestimating the growth in
BCGs between the low and high-redshift subsamples by a similar amount.
Hence, instead of finding that the mass grows by a factor of \growth,
we would
find that the mass grows by a factor of \growthadj.

In this paper, we do not use the integrated magnitude in GALFIT to
estimate stellar masses. We make this choice because most of the BCGs
in our low-redshift subsample have not been analysed with
GALFIT. 
%Secondly, the magnitude in GALFIT is very sensitive to the sky
%level.

Instead, we note that there is a source of systematic uncertainty in
the relative stellar masses between low and high-redshifts that comes
from the photometry. By comparing two widely used techniques to
do galaxy photometry, we estimate this uncertainty to be $\sim
20$\%.

\subsection{Stellar masses of the BCGs}

In section~\ref{sec:colour}, we described how we used the Ks-band
magnitude of BCGs and the predictions of a model that broadly
describes the change in the J-Ks colour of BCGs with redshift to
estimate their stellar masses. The masses will depend on the model
used, so our conclusions are model dependent. To explore how sensitive
this dependence is, we re-estimate the masses using another stellar
population synthesis code.

%We choose the two models in Figure~\ref{fig:models} that bracket the
%change in the J-Ks colour as a function of redshift (models 1 and 2)
%for this study. Model 1, which represents an instantaneous
%solar-metallicity burst at $z=2$, predicts J-Ks colours that are far
%too blue at high redshift. At the other extreme, model 2, which
%represents instantaneous burst at $z=5$ with a more metal rich
%population, predicts J-Ks colours that are far too red at high
%redshift.

%Examining the stellar mass ratio between low and high-redshift
%subsamples, we find that the stellar mass ratio varies from $0.44 \pm 0.03$
%for model 1 to $0.90 \pm 0.05$ for model 2. The stellar mass ratio for model
%3, the model we use in this paper, is $0.54 \pm 0.03$.  Clearly, adopting
%extreme models that either of these extreme models does not change the
%fundamental conclusion of the paper.

Model 5 in Fig.~\ref{fig:models} and Table \ref{tab:models} is from
\citet[hereafter M05]{Maraston2005}.  The stars in this model formed
in a single burst at $z=4$, have a metallicity that is twice solar and
form a red horizontal branch, as found in most metal-rich globular
clusters. The model follows the evolution of the $J-Ks$ colour with
redshift almost as well as model 3, although it tends to predict
redder colours at $z\sim 1.4$.

The M05 and BC03 models differ in several ways. One of the differences
most noted in the literature \citep[see][for
example]{Maraston2006,Marigo2008} is the treatment of the thermally
pulsing asymptotic giant branch (TP-AGB) phase of stellar evolution.  This
phase significantly affects the optical-NIR colours of simple stellar
populations in the age range $0.5 < t < 1.5$\,Gyr. Over this age
range, the M05 models predict redder optical-NIR colours and, for a
given stellar mass, higher NIR luminosities.

As we did for the BC03 models, we normalise the M05 model so that they
match the brightness of the BCGs over the redshift interval $0 < z <
0.3$. Using the M05 models, we find that the stellar mass of BCGs
at $z=0.9$ are $1.81 \pm0.26$ times less massive than BCGs at $z=0.2$, which
is similar to the results that we derive using the BC03 models.

We do not know if the BC03 models are more appropriate than those in
M05; however, we note that recent observations are now suggesting that
the contribution from TP-AGB stars to the near-IR flux, may not be as
significant as previously thought \citep{Kriek2010, Zibetti2012}.

Finally, for a couple of BCGs, we examine how well our stellar mass
estimates compare with measurements that are made using more extensive
photometric data, a different stellar population model and a different
way of estimating total magnitudes. Using 10 broad band filters
extending from $4640\,{\mathrm \AA}$ to $8.0\,\mu m$ (rest frame),
Rettura et al. (in preparation) estimate a stellar mass of $3.9 \times
10^{11}M_{\odot}$ for the BCG in SpARCS-J003550-431224. Our estimate
from the K band photometry is $3.3\times 10^{11}M_{\odot}$.
\citet{Rettura2006}, using 9-band photometry, derive a stellar mass of
$\sim2.3\times 10^{11}M_{\odot}$ for the BCG in
RDCS~J1252.9-2927. From the K band photometry, we find $3.3\times
10^{11}M_{\odot}$.

\subsection{Comparison with other results}\label{sec:comp}

Our finding of significant evolution in the stellar mass of BCGs
with time differs from the findings of a number of authors
\citep{Whiley2008,Collins2009,Stott2010}.  Since the Ks-band magnitudes and
J-Ks colours of the BCGs that we have added in this paper are similar to
the magnitudes and colours of BCGs from earlier works
\citep{Stott2008,Stott2010} and since much of our sample consists
of BCGs from these works, the reason for the difference lies in the
way we have done the analysis. 

In part, the difference comes from the way we have compared low and
high-redshift BCGs. In this paper, we first match clusters according
to the mass they will have by the current epoch, before comparing the
stellar mass of the BCGs they host. Earlier works have done this
comparison differently. For example, \citet{Whiley2008} match the
clusters according to the mass they had at the redshift they were
observed. We repeated our analysis using this approach (see
Table~\ref{tab:comparison2} and Sec.~\ref{sec:ClusterMasses}) and
found that the evidence for evolution became considerably weaker.

The difference may also come from the redshift interval that we use to
define the low-redshift subsample. In this paper, we use $z <
0.3$. This is broader than that used by other authors,
e.g.~\citet{Stott2008}. The broader interval allows
us to use more objects to determine the stellar mass ratio at the
expense of a smaller time interval between the low and high-redshift
samples. We repeated our analysis with the redshift interval for the
low-redshift sample set to $z<0.1$. We find that the stellar mass
ratio between the high and low-redshift subsamples increases slightly
to $0.61 \pm 0.19$. The uncertainty is larger because there are fewer objects
in the low-redshift subsample. More significant, however, is that the redshift
interval between the low and high-redshift subsamples increase,
thereby increasing the tension between the data and the predictions of
the semi-anlaytic models.

%The number of objects in our low, intermediate and high redshift
%subsamples with contemporary measurements of the cluster mass is
%adequate for detecting evolution in the stellar mass of BCGs. With
%this sample, there is some tension between the observations and the
%semi-analytic model of \cipet{DeLucia2007}. It would be interesting to
%obtain a larger sample with contemporary cluster masses to see if the
%tension increases.

\subsection{The build up of stellar mass in BCGs}

In semi-analytic models of \citep{DeLucia2007}, the stellar masses of
BCGs increase by a factor of about 3 between $z \sim 0.9$ and $z
\sim 0.2$. Our results suggest that the growth is slower than this.
Over the same redshift range, we find the increase to be a factor of \growth,
suggesting that the model over-predicts the amount of stellar mass by a
factor of $\sim 1.5$ -- the difference between the blue line and the
green triangle in Fig.~\ref{fig:massevol}).

In semi-analytic models, most of the build up in stellar mass occurs
through dry mergers (both major and minor) with other galaxies. There
is ample observational evidence for major mergers in the centres of
clusters
\citep{Rasmussen2010,Brough2011,Bildfell2012}. \citet{Brough2011} in a
study of three BCGs at $z\sim0.1$ with nearby companions found that
the companions of two of the BCGs would merge with the BCG within
0.35\,Gyr. More dramatic still is the the merger that is occurring in
the centre of MZ 10451 \citep{Rasmussen2010}.

Evidence for major mergers can also be seen in some of the SpARCS BCGs, For
example, the isophotes of the BCG in SpARCS~J163435+402151 are
distorted, indicating a possible major merger. In this cluster, there
is evidence that another major merger is occurring for a galaxy that
is almost as bright on the other side of the cluster.

There is also an example in the SpARCS sample of a merger that is
likely to happen by today. In the centre of SpARCS~J161641+554513,
there is a galaxy that is within 20\,kpc projected distance of the
BCG. The velocity difference between the two galaxies is $\sim
140$\,km/s, and the companion is almost as bright as the BCG (see
Fig.~\ref{fig:BCGimages}). It is highly likely that these two galaxies
would have merged by now. Additional examples of likely major mergers
can be found in RX~J0848.9+4452 \citep{Yamada2002} and
RDCS~J1252-2927 \citep{Collins2009}.

While it is clear that mergers do occur, it is not yet clear what
fraction of the stars in the merging galaxies end up in the BCG and
what fraction end up distributed throughout the cluster, appearing as
intra-cluster light (ICL). The apparent lack of evolution in the
stellar mass of BCGs that was found in earlier work suggested that the
contribution to the ICL was close to 100\%
\citep{Whiley2008,Collins2009,Stott2010}. Our results suggest that it
is closer to 50\%.  High resolution simulations suggest that 50-80\% of
the mass of mergers will be distributed throughout the cluster
\citep{Conroy2007,Puchwein2010}.  Recent measurements of the ICL
show that the ICL grows relative to the total cluster light by a factor of
2--4 since $z \sim 1$ \citep{Burke2012}.
%Direct observational evidence for
%this can be seen in the merger that is occurring in the centre of MZ
%10451 \citep{Rasmussen2010}.

It is possible that some of the BCGs in our sample are increasing
their stellar mass through star formation. Out of the 12 SpARCS BCGs,
5 show emission from the [OII]\,$\lambda\lambda$\,3726,3728 doublet,
which is an indicator of star formation and/or AGN activity. It is
unlikely that most of the [OII] emission that we detect comes from
star formation. Over 70\% of low redshift BCGs that have detectable
[OII] emission have line ratios that are consistent with the line
ratios of AGN \citep{vonderLinden2007}. Our spectra do not cover the
lines that can be used to separate between AGN activity and star
formation, such as the [OIII]\,$\lambda\lambda$\,4959,5007 doublet,
H${\beta}$, H$\alpha$ and [NII]$\lambda$\,6584.

If we were to assume that the [OII] emission did come from star
formation entirely and if we ignore dust, then the average [OII] line
flux corresponds to a star formation rate of about 1 solar mass per
year, using the equation (4) in \citet{Kewley2004} to make the
conversion between [OII] line flux and the star formation rate. At
these rates, star-formation will not contribute much to the overall
stellar mass of the BCG, even if they were to continue forming stars
at this rate until today.

%\citep{DeLucia2007} note that the main progenitors of low redshift
%BCGs are not always going to be BCGs at earlier
%epochs. Observationally, this means that one might select a galaxy in
%a high redshift cluster that is not destined to be the BCG at lower
%redshifts, even though the galaxy is the BCG in the cluster at the
%epoch the cluster was observed. Therefore, in the data, one is generally not
%comparing local BCGs with their main progenitors at high redshift, but
%local BCGs with distant ones. Compared to the models, the stellar mass
%ratio that is measured from observations will therefore be biased high.

\section{Summary and Conclusions}

Using near-IR photometry from the literature
\citep{Stott2008,Stott2010} and photometry from an analysis of imaging
data that we obtained using several ground-based near-IR cameras, we
have investigated how the stellar masses of BCGs change with
redshift. The BCGs in our sample cover a broad redshift range, from
$z=0.03$ to $z=1.63$, which covers 9.8\,Gyr, or 70\% of the
history of the universe.

To estimate the stellar mass of the BCGs, we compare the Ks band flux with the
predictions from a stellar population synthesis model that matches the
J-Ks colour of the BCGs over the entire redshift range covered by the data.

We then compare mass of BCGs at low and high redshifts.  After
accounting for the correlation between BCG stellar mass and cluster
mass, we find that, between $z=0.9$ and $z=0.2$, BCGs, on average,
grow in mass by a factor of \growth. Our result is not weakened if we
choose other methods to estimate the Ks band flux or if we choose
other stellar population synthesis models to infer the mass. The
systematic uncertainty coming from the photometry is probably the
dominant source of systematic uncertainty in our analysis and affects
our estimates of the growth rate by around 20\%.

Our conclusions differ from those of earlier works
\citep{Collins2009,Stott2010}. In part, this is due to the way we have
accounted for the correlation between the mass of the BCG and the mass
of the cluster and to the redshift intervals that we use to define the
low and high-redshift subsamples.

Our measurements are now in better agreement with the predictions of
semi-analytic models for the growth of stellar mass in BCGs
\citep{DeLucia2007}. However, there is still some tension between the
data and these models, which predict growth rates that are a factor of
1.5 higher.

We find direct evidence that some of the BCGs in our sample are
building up their stellar mass through star-formation and major
mergers. However, star-formation, while present in some of BCGs, is at
low levels. At these levels, star formation cannot be the dominant
mechanism for the build up of stellar mass in BCGs over the last 10
billion years. The build-up mainly occurs through mergers, of which
some are clearly major.

%Presumably, minor mergers also occur. Indeed, some semi-analytic
%models suggest that this is the dominant mechanism.

% For future papers

%\item Construct the J-Ks colour magnitude diagram highlighting the location of spectroscopically confirmed cluster members.

%\item Use the spectroscopically confirmed cluster members to define the red sequence

%\item Compare the colour of the BCG with respect to the line defined by the red sequence. Is it bluer/redder?

%\end{itemize}

\section*{Acknowledgments}

The authors thank John Stott, Chris Collins and Gabriela DeLucia for
providing us with the tabulated data from their papers and for useful
discussions. The data in this paper were based in part on observations
obtained with WIRCam, a joint project of CFHT, Taiwan, Korea, Canada,
France, at the Canada-France-Hawaii Telescope (CFHT) which is operated
by the National Research Council (NRC) of Canada, the Institute
National des Sciences de l'Univers of the Centre National de la
Recherche Scientifique of France, and the University of Hawaii. Based
in part on observations taken at the ESO Paranal Observatory (ESO
programmes 084.A-0214, 085.A-0166, and 085.A-0613). Based in part on observations
taken at the Cerro Tololo Inter-American Observatory. R.D. gratefully
acknowledges the support provided by the BASAL Center for Astrophysics
and Associated Technologies (CATA), and by FONDECYT grant N. 1100540.
C.L. is the recipient of an Australian Research Council Future
Fellowship (program number FT0992259). G.W. gratefully acknowledges
support from NSF grant AST-0909198.

% Add a phrase which respects the traditional owners of the land.

%\bibliography{IR}
\appendix

\section{Data}

In Tables~\ref{tab:app1}, \ref{tab:app2} and \ref{tab:app3}, we list
the names of the clusters, their redshifts, their masses and the mass
proxy used to determine masses. Two estimates of the mass are
provided. The first is computed from the mass proxy and represents the
mass of the cluster when it was observed.The errors in the
 cluster masses only include the error in the mass proxy. They do not
 include the intrinsic scatter in the relation between mass and
 mass proxy.  The second mass extrapolates the
first mass to the current  epoch by integrating the mean
 mass accretion rates in \citet{Fakhouri2010}.  The error does not
 take into account the intrinsic scatter
in the accretion rates. \citet{Wechsler2002} estimates that between $z=1$ and $z=0$,
the final mass of a $10^{14}\,M_{\odot}$ halo can scatter by 20-30\%. Also
listed in the table are the magnitudes and colours of the BCGs. If
available, we also list the errors in these quantities.
  Excluding the BCGs in the SpARCS and CNOC1 clusters, the magnitudes
  and colours of the BCGs in these tables were obtained from
  \citet{Stott2008,Stott2010} and J.~P.~Stott (private
  communication). The model-dependent masses, which have been
adjusted to account for the loss of mass due to supernova explosions
and stellar winds, are listed in the final column. Not all clusters
have mass measurements in the papers listed in the main body of the
paper. These clusters are listed in Table~\ref{tab:nomass}.

\begin{table*}
\caption{The low-redshift subsample}\label{tab:app1}
\centering           
\begin{tabular}{lrrrlllr}
\hline
\multicolumn{5}{c}{Cluster} & \multicolumn{3}{c}{BCG}\\
\hline 
Name  & Redshift  & Mass & Mass today & Mass Proxy & Ks  &  J-Ks  &
Stellar Mass \\
      &                & [$10^{15}\,M_{\odot}$] &
      [$10^{15}\,M_{\odot}$] & & [mag] & [mag] &
      [$10^{12}\,M_{\odot}$] \\
\hline
Abell1902                 & 0.160 & $0.48^{+0.06}_{-0.06}$ & $0.61^{+0.08}_{-0.08}$ & X--ray luminosity & $12.63$ & $1.37$ &  0.59  \\[2pt]
Abell193                  & 0.049 & $0.18^{+0.03}_{-0.03}$ & $0.20^{+0.03}_{-0.03}$ & X--ray luminosity & $10.43$ & $1.06$ &  0.48  \\[2pt]
Abell1930                 & 0.131 & $0.38^{+0.05}_{-0.05}$ & $0.48^{+0.07}_{-0.07}$ & X--ray luminosity & $12.47$ & $1.10$ &  0.48  \\[2pt]
Abell1991                 & 0.059 & $0.16^{+0.02}_{-0.02}$ & $0.18^{+0.02}_{-0.02}$ & X--ray luminosity & $11.15$ & $1.01$ &  0.35  \\[2pt]
Abell2029                 & 0.077 & $ 1.2^{+0.2}_{-0.2}$ & $ 1.4^{+0.2}_{-0.2}$ &  X-ray gas mass & $10.30$ & $1.13$ &  1.31  \\[2pt]
Abell2034                 & 0.113 & $0.89^{+0.13}_{-0.13}$ & $ 1.2^{+0.2}_{-0.2}$ &  X-ray gas mass & $12.21$ & $1.07$ &  0.46  \\[2pt]
Abell2052                 & 0.035 & $0.25^{+0.02}_{-0.02}$ & $0.28^{+0.02}_{-0.02}$ & X--ray luminosity & $ 9.88$ & $1.00$ &  0.41  \\[2pt]
Abell2065                 & 0.073 & $0.43^{+0.04}_{-0.04}$ & $0.48^{+0.05}_{-0.05}$ & X--ray luminosity & $12.03$ & $1.15$ &  0.24  \\[2pt]
Abell2072                 & 0.127 & $0.31^{+0.06}_{-0.06}$ & $0.39^{+0.08}_{-0.08}$ & X--ray luminosity & $12.82$ & $1.23$ &  0.33  \\[2pt]
Abell2107                 & 0.041 & $0.13^{+0.02}_{-0.02}$ & $0.15^{+0.02}_{-0.02}$ & X--ray luminosity & $10.10$ & $1.00$ &  0.45  \\[2pt]
Abell2124                 & 0.066 & $0.16^{+0.03}_{-0.03}$ & $0.17^{+0.03}_{-0.03}$ & X--ray luminosity & $11.05$ & $1.05$ &  0.48  \\[2pt]
Abell2175                 & 0.095 & $0.29^{+0.03}_{-0.03}$ & $0.32^{+0.04}_{-0.04}$ & X--ray luminosity & $11.78$ & $1.18$ &  0.50  \\[2pt]
Abell2204                 & 0.152 & $ 1.4^{+0.2}_{-0.2}$ & $ 1.8^{+0.3}_{-0.3}$ &  X-ray gas mass & $12.23$ & $1.14$ &  0.78  \\[2pt]
Abell2244                 & 0.097 & $0.82^{+0.15}_{-0.15}$ & $0.94^{+0.17}_{-0.17}$ &  X-ray gas mass & $11.59$ & $1.11$ &  0.62  \\[2pt]
Abell2259                 & 0.164 & $0.55^{+0.08}_{-0.08}$ & $0.71^{+0.10}_{-0.10}$ & X--ray luminosity & $12.77$ & $1.19$ &  0.54  \\[2pt]
Abell2345                 & 0.177 & $0.76^{+0.13}_{-0.13}$ & $0.98^{+0.17}_{-0.17}$ & X--ray luminosity & $12.60$ & $1.24$ &  0.72  \\[2pt]
Abell2377                 & 0.081 & $0.31^{+0.05}_{-0.05}$ & $0.34^{+0.06}_{-0.06}$ & X--ray luminosity & $12.07$ & $1.09$ &  0.28  \\[2pt]
Abell2382                 & 0.062 & $0.12^{+0.03}_{-0.03}$ & $0.13^{+0.03}_{-0.03}$ & X--ray luminosity & $11.43$ & $1.06$ &  0.30  \\[2pt]
Abell2384                 & 0.094 & $0.55^{+0.06}_{-0.06}$ & $0.63^{+0.07}_{-0.07}$ & X--ray luminosity & $12.59$ & $1.15$ &  0.23  \\[2pt]
Abell2402                 & 0.081 & $0.22^{+0.04}_{-0.04}$ & $0.24^{+0.05}_{-0.05}$ & X--ray luminosity & $11.67$ & $1.17$ &  0.41  \\[2pt]
Abell2415                 & 0.058 & $0.19^{+0.03}_{-0.03}$ & $0.21^{+0.03}_{-0.03}$ & X--ray luminosity & $11.46$ & $1.05$ &  0.26  \\[2pt]
Abell2426                 & 0.098 & $0.44^{+0.07}_{-0.07}$ & $0.50^{+0.08}_{-0.08}$ & X--ray luminosity & $12.07$ & $1.02$ &  0.41  \\[2pt]
Abell2428                 & 0.085 & $0.25^{+0.05}_{-0.05}$ & $0.28^{+0.05}_{-0.05}$ & X--ray luminosity & $11.64$ & $1.07$ &  0.46  \\[2pt]
Abell2443                 & 0.108 & $0.31^{+0.05}_{-0.05}$ & $0.39^{+0.06}_{-0.06}$ & X--ray luminosity & $11.98$ & $1.13$ &  0.53  \\[2pt]
Abell2457                 & 0.059 & $0.16^{+0.04}_{-0.04}$ & $0.18^{+0.04}_{-0.04}$ & X--ray luminosity & $10.83$ & $1.03$ &  0.48  \\[2pt]
Abell2495                 & 0.078 & $0.29^{+0.04}_{-0.04}$ & $0.33^{+0.04}_{-0.04}$ & X--ray luminosity & $11.69$ & $1.09$ &  0.37  \\[2pt]
Abell2496                 & 0.123 & $0.35^{+0.11}_{-0.11}$ & $0.44^{+0.14}_{-0.14}$ & X--ray luminosity & $11.92$ & $1.17$ &  0.71  \\[2pt]
Abell2589                 & 0.042 & $0.20^{+0.02}_{-0.02}$ & $0.22^{+0.02}_{-0.02}$ & X--ray luminosity & $10.31$ & $1.04$ &  0.39  \\[2pt]
Abell2593                 & 0.043 & $0.14^{+0.02}_{-0.02}$ & $0.16^{+0.02}_{-0.02}$ & X--ray luminosity & $10.36$ & $1.02$ &  0.39  \\[2pt]
Abell2597                 & 0.085 & $0.38^{+0.07}_{-0.07}$ & $0.43^{+0.08}_{-0.08}$ &  X-ray gas mass & $12.31$ & $1.01$ &  0.25  \\[2pt]
Abell2622                 & 0.062 & $0.13^{+0.02}_{-0.02}$ & $0.15^{+0.02}_{-0.02}$ & X--ray luminosity & $11.38$ & $1.01$ &  0.32  \\[2pt]
Abell2626                 & 0.057 & $0.21^{+0.02}_{-0.02}$ & $0.23^{+0.02}_{-0.02}$ & X--ray luminosity & $10.75$ & $1.09$ &  0.48  \\[2pt]
Abell2627                 & 0.126 & $0.32^{+0.06}_{-0.06}$ & $0.41^{+0.08}_{-0.08}$ & X--ray luminosity & $12.51$ & $1.22$ &  0.43  \\[2pt]
Abell2717                 & 0.050 & $0.12^{+0.02}_{-0.02}$ & $0.14^{+0.02}_{-0.02}$ & X--ray luminosity & $10.94$ & $1.06$ &  0.31  \\[2pt]
Abell2734                 & 0.062 & $0.26^{+0.03}_{-0.03}$ & $0.29^{+0.03}_{-0.03}$ & X--ray luminosity & $11.17$ & $1.05$ &  0.38  \\[2pt]
Abell376                  & 0.049 & $0.16^{+0.02}_{-0.02}$ & $0.18^{+0.02}_{-0.02}$ & X--ray luminosity & $10.78$ & $1.11$ &  0.35  \\[2pt]
Abell399                  & 0.072 & $0.52^{+0.05}_{-0.05}$ & $0.59^{+0.06}_{-0.06}$ & X--ray luminosity & $10.84$ & $1.01$ &  0.70  \\[2pt]
Abell401                  & 0.074 & $ 1.3^{+0.2}_{-0.2}$ & $ 1.5^{+0.2}_{-0.2}$ &  X-ray gas mass & $10.91$ & $1.22$ &  0.69  \\[2pt]
Abell115                  & 0.197 & $ 1.0^{+0.2}_{-0.2}$ & $ 1.3^{+0.2}_{-0.2}$ & X--ray luminosity & $13.40$ & $1.25$ &  0.41  \\[2pt]
Abell1201                 & 0.169 & $0.53^{+0.09}_{-0.09}$ & $0.68^{+0.12}_{-0.12}$ & X--ray luminosity & $13.16$ & $1.29$ &  0.40  \\[2pt]
Abell1204                 & 0.171 & $0.59^{+0.09}_{-0.09}$ & $0.76^{+0.12}_{-0.12}$ & X--ray luminosity & $13.58$ & $1.12$ &  0.28  \\[2pt]
Abell1246                 & 0.190 & $0.62^{+0.10}_{-0.10}$ & $0.79^{+0.13}_{-0.13}$ & X--ray luminosity & $13.67$ & $1.32$ &  0.30  \\[2pt]
Abell1423                 & 0.213 & $ 1.2^{+0.3}_{-0.3}$ & $ 1.8^{+0.4}_{-0.4}$ &  X-ray gas mass & $13.65$ & $1.41$ &  0.38  \\[2pt]
Abell1553                 & 0.165 & $0.59^{+0.09}_{-0.09}$ & $0.76^{+0.11}_{-0.11}$ & X--ray luminosity & $12.60$ & $1.11$ &  0.64  \\[2pt]
Abell1682                 & 0.234 & $ 1.7^{+0.4}_{-0.4}$ & $ 2.5^{+0.7}_{-0.7}$ &  X-ray gas mass & $13.12$ & $1.31$ &  0.72  \\[2pt]

\hline
\multicolumn{8}{r}{\it Continued on next page ...}
\end{tabular}
\end{table*}

\setcounter{table}{0}

\begin{table*}
\caption{\it --- continued from previous page}\label{tab:app1}
\centering           
\begin{tabular}{lrlllllr}
\hline
\multicolumn{5}{c}{Cluster} & \multicolumn{3}{c}{BCG}\\
\hline 
Name  & Redshift  & Mass & Mass today & Mass Proxy & Ks  &  J-Ks  &
Stellar Mass \\
      &                & [$10^{15}\,M_{\odot}$] &
      [$10^{15}\,M_{\odot}$] & & [mag] & [mag] &
      [$10^{12}\,M_{\odot}$] \\
\hline
Abell1704                 & 0.221 & $0.63^{+0.11}_{-0.11}$ & $0.92^{+0.17}_{-0.17}$ & X--ray luminosity & $13.56$ & $1.23$ &  0.43  \\[2pt]
Abell1758                 & 0.279 & $0.87^{+0.14}_{-0.14}$ & $ 1.3^{+0.2}_{-0.2}$ & X--ray luminosity & $13.96$ & $1.38$ &  0.45  \\[2pt]
Abell1763                 & 0.223 & $ 2.3^{+0.5}_{-0.5}$ & $ 3.5^{+0.7}_{-0.7}$ &  X-ray gas mass & $13.11$ & $1.33$ &  0.67  \\[2pt]
Abell1835                 & 0.253 & $ 1.7^{+0.2}_{-0.2}$ & $ 2.5^{+0.3}_{-0.3}$ &  X-ray gas mass & $12.92$ & $1.44$ &  0.99  \\[2pt]
Abell1914                 & 0.171 & $ 1.4^{+0.2}_{-0.2}$ & $ 1.9^{+0.3}_{-0.3}$ &  X-ray gas mass & $12.85$ & $1.24$ &  0.54  \\[2pt]
Abell1961                 & 0.232 & $0.55^{+0.10}_{-0.10}$ & $0.81^{+0.14}_{-0.14}$ & X--ray luminosity & $13.52$ & $1.34$ &  0.49  \\[2pt]
Abell2009                 & 0.153 & $0.70^{+0.10}_{-0.10}$ & $0.91^{+0.13}_{-0.13}$ & X--ray luminosity & $12.83$ & $1.19$ &  0.46  \\[2pt]
Abell209                  & 0.209 & $ 1.7^{+0.3}_{-0.3}$ & $ 2.6^{+0.4}_{-0.4}$ &  X-ray gas mass & $13.07$ & $1.39$ &  0.62  \\[2pt]
Abell2111                 & 0.229 & $ 1.1^{+0.2}_{-0.2}$ & $ 1.7^{+0.4}_{-0.4}$ &  X-ray gas mass & $13.75$ & $1.35$ &  0.39  \\[2pt]
Abell2163                 & 0.203 & $ 5.2^{+0.7}_{-0.7}$ & $ 8.2^{+1.1}_{-1.1}$ &  X-ray gas mass & $13.24$ & $1.74$ &  0.51  \\[2pt]
Abell2218                 & 0.176 & $0.96^{+0.16}_{-0.16}$ & $ 1.3^{+0.2}_{-0.2}$ &  X-ray gas mass & $13.35$ & $1.11$ &  0.36  \\[2pt]
Abell2219                 & 0.226 & $ 2.5^{+0.3}_{-0.3}$ & $ 3.9^{+0.5}_{-0.5}$ &  X-ray gas mass & $13.34$ & $1.36$ &  0.55  \\[2pt]
Abell2254                 & 0.178 & $0.62^{+0.09}_{-0.09}$ & $0.80^{+0.12}_{-0.12}$ & X--ray luminosity & $13.19$ & $1.27$ &  0.42  \\[2pt]
Abell2261                 & 0.224 & $ 1.9^{+0.4}_{-0.4}$ & $ 3.0^{+0.6}_{-0.6}$ &  X-ray gas mass & $12.62$ & $1.42$ &  1.06  \\[2pt]
Abell2445                 & 0.165 & $0.37^{+0.08}_{-0.08}$ & $0.47^{+0.10}_{-0.10}$ & X--ray luminosity & $13.22$ & $1.18$ &  0.36  \\[2pt]
Abell2561                 & 0.163 & $0.32^{+0.08}_{-0.08}$ & $0.40^{+0.10}_{-0.10}$ & X--ray luminosity & $13.53$ & $1.18$ &  0.27  \\[2pt]
Abell521                  & 0.248 & $ 1.5^{+0.2}_{-0.2}$ & $ 2.3^{+0.4}_{-0.4}$ &  X-ray gas mass & $13.53$ & $1.33$ &  0.55  \\[2pt]
Abell586                  & 0.171 & $0.82^{+0.12}_{-0.12}$ & $ 1.1^{+0.2}_{-0.2}$ & X--ray luminosity & $13.13$ & $1.23$ &  0.42  \\[2pt]
Abell661                  & 0.288 & $0.98^{+0.21}_{-0.21}$ & $ 1.5^{+0.3}_{-0.3}$ & X--ray luminosity & $13.53$ & $1.31$ &  0.71  \\[2pt]
Abell665                  & 0.182 & $ 1.7^{+0.2}_{-0.2}$ & $ 2.2^{+0.3}_{-0.3}$ &  X-ray gas mass & $13.69$ & $1.23$ &  0.28  \\[2pt]
Abell68                   & 0.255 & $ 1.0^{+0.2}_{-0.2}$ & $ 1.5^{+0.3}_{-0.2}$ &  X-ray gas mass & $13.51$ & $1.43$ &  0.58  \\[2pt]
Abell750                  & 0.180 & $0.72^{+0.12}_{-0.12}$ & $0.93^{+0.15}_{-0.15}$ & X--ray luminosity & $13.05$ & $1.31$ &  0.49  \\[2pt]
Abell773                  & 0.217 & $ 1.2^{+0.1}_{-0.1}$ & $ 1.7^{+0.2}_{-0.2}$ &  X-ray gas mass & $13.22$ & $1.42$ &  0.58  \\[2pt]
Abell907                  & 0.153 & $0.63^{+0.09}_{-0.09}$ & $0.81^{+0.12}_{-0.12}$ & X--ray luminosity & $13.18$ & $1.32$ &  0.33  \\[2pt]
Abell963                  & 0.206 & $0.91^{+0.13}_{-0.13}$ & $ 1.4^{+0.2}_{-0.2}$ &  X-ray gas mass & $12.94$ & $1.37$ &  0.68  \\[2pt]
RX J1720.1+2638            & 0.164 & $ 1.1^{+0.1}_{-0.1}$ & $ 1.4^{+0.1}_{-0.1}$ & X--ray luminosity & $12.97$ & $1.21$ &  0.45  \\[2pt]
RX J2129.6+0005            & 0.235 & $ 1.0^{+0.2}_{-0.2}$ & $ 1.5^{+0.3}_{-0.2}$ &  X-ray gas mass & $13.27$ & $1.37$ &  0.63  \\[2pt]
Zw1432                    & 0.186 & $0.46^{+0.11}_{-0.11}$ & $0.59^{+0.14}_{-0.14}$ & X--ray luminosity & $13.31$ & $1.31$ &  0.41  \\[2pt]
Zw1693                    & 0.225 & $0.61^{+0.15}_{-0.15}$ & $0.89^{+0.23}_{-0.23}$ & X--ray luminosity & $13.24$ & $1.38$ &  0.60  \\[2pt]
Zw1883                    & 0.194 & $0.54^{+0.13}_{-0.13}$ & $0.69^{+0.17}_{-0.17}$ & X--ray luminosity & $13.22$ & $1.28$ &  0.48  \\[2pt]
Zw2089                    & 0.230 & $0.42^{+0.05}_{-0.05}$ & $0.60^{+0.08}_{-0.08}$ &  X-ray gas mass & $14.10$ & $1.40$ &  0.28  \\[2pt]
Zw2379                    & 0.205 & $0.49^{+0.10}_{-0.10}$ & $0.71^{+0.15}_{-0.15}$ & X--ray luminosity & $13.86$ & $1.27$ &  0.29  \\[2pt]
Zw2701                    & 0.214 & $0.54^{+0.09}_{-0.09}$ & $0.78^{+0.14}_{-0.14}$ &  X-ray gas mass & $13.40$ & $1.30$ &  0.48  \\[2pt]
Zw348                     & 0.255 & $0.76^{+0.16}_{-0.16}$ & $ 1.1^{+0.2}_{-0.2}$ & X--ray luminosity & $13.78$ & $1.43$ &  0.45  \\[2pt]
Zw3916                    & 0.206 & $0.54^{+0.08}_{-0.08}$ & $0.78^{+0.12}_{-0.12}$ & X--ray luminosity & $13.80$ & $1.31$ &  0.31  \\[2pt]
Zw5247                    & 0.195 & $ 1.1^{+0.3}_{-0.3}$ & $ 1.4^{+0.3}_{-0.3}$ &  X-ray gas mass & $13.41$ & $1.28$ &  0.40  \\[2pt]
Zw5768                    & 0.266 & $0.86^{+0.14}_{-0.14}$ & $ 1.3^{+0.2}_{-0.2}$ & X--ray luminosity & $13.08$ & $1.25$ &  0.93  \\[2pt]
Zw7215                    & 0.292 & $0.84^{+0.17}_{-0.17}$ & $ 1.2^{+0.3}_{-0.3}$ & X--ray luminosity & $14.12$ & $1.45$ &  0.43  \\[2pt]
Abell2390                     & 0.228 & $ 3.2^{+0.6}_{-0.5}$ & $ 5.0^{+0.9}_{-0.8}$ & X--ray temperature & $13.49 \pm 0.07$ & ... &  0.49  \\[2pt]
MS0440+02                 & 0.197 & $ 1.4^{+0.8}_{-0.3}$ & $ 1.8^{+1.1}_{-0.4}$ & X--ray temperature & $13.34 \pm 0.05$ & ... &  0.44  \\[2pt]
MS0451+02                 & 0.201 & $0.78^{+0.20}_{-0.15}$ & $ 1.2^{+0.3}_{-0.2}$ & X--ray temperature & $13.94 \pm 0.07$ & ... &  0.26  \\[2pt]
MS0839+29                 & 0.193 & $0.32^{+0.05}_{-0.05}$ & $0.41^{+0.07}_{-0.06}$ & X--ray temperature & $13.41 \pm 0.06$ & ... &  0.40  \\[2pt]
MS1006+12                 & 0.261 & $ 1.0^{+0.3}_{-0.3}$ & $ 1.5^{+0.5}_{-0.4}$ & X--ray temperature & $13.79 \pm 0.07$ & ... &  0.47  \\[2pt]
MS1231+15                 & 0.235 & $0.46^{+0.14}_{-0.14}$ & $0.66^{+0.21}_{-0.21}$ & Velocity Dispersion & $13.89 \pm 0.07$ & ... &  0.36  \\[2pt]
MS1455+22                 & 0.257 & $0.71^{+0.25}_{-0.16}$ & $ 1.0^{+0.4}_{-0.2}$ & X--ray temperature & $13.56 \pm 0.06$ & ... &  0.57  \\[2pt]

\hline
\end{tabular}
\end{table*}

\begin{table*}
\caption{The intermediate-redshift subsample}\label{tab:app2}
\centering           
\begin{tabular}{lrlllllr}
\hline
\multicolumn{5}{c}{Cluster} & \multicolumn{3}{c}{BCG}\\
\hline 
Name  & Redshift  & Mass & Mass today & Mass Proxy & Ks  &  J-Ks  &
Stellar Mass \\
      &                & [$10^{15}\,M_{\odot}$] &
      [$10^{15}\,M_{\odot}$] & & [mag] & [mag] &
      [$10^{12}\,M_{\odot}$] \\
\hline
MACS J0018.5+1626          & 0.541 & $ 1.7^{+0.5}_{ 0.5}$ & $ 3.9^{+1.3}_{ 1.3}$ & X--ray temperature & $15.35$ & $1.52$ &  0.50  \\[2pt]
MACS J0025.4-1222          & 0.478 & $0.95^{+0.21}_{ 0.21}$ & $ 1.9^{+0.4}_{ 0.4}$ & X--ray temperature & $15.70$ & $1.67$ &  0.28  \\[2pt]
MACS J0257.6-2209          & 0.504 & $ 2.1^{+0.4}_{ 0.4}$ & $ 5.0^{+1.2}_{ 1.2}$ & X--ray temperature & $14.77$ & $1.61$ &  0.74  \\[2pt]
MACS J0404.6+1109          & 0.358 & $ 1.2^{+1.1}_{ 1.1}$ & $ 2.0^{+2.0}_{ 2.0}$ & X--ray temperature & $13.82$ & ... &  0.85  \\[2pt]
MACS J0429.6-0253          & 0.397 & $ 1.4^{+0.6}_{ 0.6}$ & $ 2.4^{+1.1}_{ 1.1}$ & X--ray temperature & $13.58$ & ... &  1.34  \\[2pt]
MACS J0454.1-0300          & 0.550 & $ 1.0^{+0.3}_{ 0.3}$ & $ 2.3^{+0.8}_{ 0.8}$ & X--ray temperature & $15.29$ & $1.58$ &  0.54  \\[2pt]
MACS J0647.7+7015          & 0.584 & $ 2.5^{+0.5}_{ 0.5}$ & $ 6.0^{+1.2}_{ 1.2}$ & X--ray temperature & $14.87$ & $1.76$ &  0.89  \\[2pt]
MACS J0744.8+3927          & 0.686 & $ 1.1^{+0.2}_{ 0.2}$ & $ 3.0^{+0.5}_{ 0.5}$ & X--ray temperature & $15.33$ & $1.88$ &  0.76  \\[2pt]
MACS J2129.4-0741          & 0.570 & $ 1.2^{+0.2}_{ 0.2}$ & $ 2.7^{+0.6}_{ 0.6}$ & X--ray temperature & $15.57$ & $1.72$ &  0.45  \\[2pt]
MACS J2245.0+2637          & 0.301 & $0.60^{+0.14}_{ 0.14}$ & $1.00^{+0.25}_{ 0.25}$ & X--ray temperature & $14.03$ & ... &  0.49  \\[2pt]
MS0016+16                 & 0.547 & $ 1.6^{+0.4}_{-0.3}$ & $ 3.7^{+1.0}_{-0.8}$ & X--ray temperature & $15.29 \pm 0.08$ & ... &  0.54  \\[2pt]
MS0302+16                 & 0.425 & $0.29^{+0.65}_{-0.13}$ & $0.53^{+1.30}_{-0.25}$ & X--ray temperature & $15.01 \pm 0.07$ & ... &  0.42  \\[2pt]
MS0451-03                 & 0.539 & $ 2.0^{+0.4}_{-0.4}$ & $ 4.6^{+1.1}_{-1.0}$ & X--ray temperature & $15.18 \pm 0.07$ & ... &  0.58  \\[2pt]
MS1008-12                 & 0.306 & $0.70^{+0.24}_{-0.16}$ & $ 1.2^{+0.4}_{-0.3}$ & X--ray temperature & $13.68 \pm 0.08$ & ... &  0.70  \\[2pt]
MS1224+20                 & 0.326 & $0.75^{+0.25}_{-0.25}$ & $ 1.3^{+0.4}_{-0.4}$ & Velocity Dispersion & $14.41 \pm 0.08$ & ... &  0.41  \\[2pt]
MS1358+62                 & 0.329 & $ 1.7^{+0.9}_{-0.4}$ & $ 2.9^{+1.6}_{-0.7}$ & X--ray temperature & $14.29 \pm 0.06$ & ... &  0.46  \\[2pt]
MS1512+36                 & 0.373 & $0.21^{+0.12}_{-0.08}$ & $0.34^{+0.20}_{-0.13}$ & X--ray temperature & $14.63 \pm 0.08$ & ... &  0.44  \\[2pt]
MS1621+26                 & 0.427 & $0.94^{+1.04}_{-0.47}$ & $ 1.8^{+2.2}_{-1.0}$ & X--ray temperature & $14.98 \pm 0.07$ & ... &  0.44  \\[2pt]

\hline
\end{tabular}
\end{table*}

\begin{table*}
\caption{The high-redshift subsample}\label{tab:app3}
\centering           
\begin{tabular}{lrlllllr}
\hline
\multicolumn{5}{c}{Cluster} & \multicolumn{3}{c}{BCG}\\
\hline 
Name  & Redshift  & Mass & Mass today & Mass Proxy & Ks  &  J-Ks  &
Stellar Mass \\
     &                & [$10^{15}\,M_{\odot}$] &
     [$10^{15}\,M_{\odot}$] & & [mag] & [mag] &
     [$10^{12}\,M_{\odot}$] \\
\hline
SpARCS J003442-430752               & 0.867 & $0.36^{+0.16}_{-0.19}$ & $ 1.1^{+0.6}_{-0.6}$ & Velocity Dispersion & $16.52 \pm 0.04$ & $1.86 \pm 0.03$ &  0.38  \\[2pt]
SpARCS J003645-441050               & 0.867 & $0.45^{+0.16}_{-0.14}$ & $ 1.4^{+0.6}_{-0.5}$ & Velocity Dispersion & $16.09 \pm 0.05$ & $1.84 \pm 0.03$ &  0.56  \\[2pt]
SpARCS J161314+564930               & 0.873 & $ 2.6^{+0.6}_{-0.5}$ & $10^{+3}_{-2}$ & Velocity Dispersion & $15.69 \pm 0.01$ & $1.79 \pm 0.01$ &  0.81  \\[2pt]
SpARCS J104737+574137               & 0.956 & $0.29^{+0.10}_{-0.13}$ & $1.00^{+0.40}_{-0.49}$ & Velocity Dispersion & $17.14 \pm 0.03$ & $1.89 \pm 0.03$ &  0.24  \\[2pt]
SpARCS J021524-034331               & 1.004 & $0.26^{+0.17}_{-0.13}$ & $ 1.0^{+0.8}_{-0.5}$ & Velocity Dispersion & $16.88 \pm 0.14$ & $1.86 \pm 0.05$ &  0.33  \\[2pt]
SpARCS J105111+581803               & 1.035 & $0.12^{+0.03}_{-0.06}$ & $0.42^{+0.12}_{-0.23}$ & Velocity Dispersion & $16.88 \pm 0.05$ & $1.74 \pm 0.03$ &  0.35  \\[2pt]
SpARCS J161641+554513               & 1.156 & $0.28^{+0.11}_{-0.12}$ & $ 1.3^{+0.6}_{-0.6}$ & Velocity Dispersion & $17.02 \pm 0.03$ & $1.73 \pm 0.02$ &  0.36  \\[2pt]
SpARCS J163435+402151               & 1.177 & $0.44^{+0.11}_{-0.16}$ & $ 2.1^{+0.6}_{-0.9}$ & Velocity Dispersion & $17.35 \pm 0.02$ & $1.84 \pm 0.03$ &  0.27  \\[2pt]
SpARCS J163852+403843               & 1.196 & $0.10^{+0.03}_{-0.05}$ & $0.38^{+0.15}_{-0.21}$ & Velocity Dispersion & $17.65 \pm 0.05$ & $1.91 \pm 0.05$ &  0.21  \\[2pt]
SpARCS J003550-431224               & 1.340 & $0.39^{+0.13}_{-0.15}$ & $ 2.5^{+1.0}_{-1.1}$ & Velocity Dispersion & $17.52 \pm 0.01$ & $1.98 \pm 0.01$ &  0.29  \\[2pt]
SpARCS J033056-284300               & 1.620 & $0.24^{+0.10}_{-0.15}$ & $ 2.1^{+1.1}_{-1.4}$ & Velocity Dispersion & $17.88 \pm 0.04$ & ... &  0.29  \\[2pt]
SpARCS J022426-032331               & 1.630 & $0.04^{+0.01}_{-0.03}$ & $0.26^{+0.05}_{-0.17}$ & Velocity Dispersion & $18.07 \pm 0.03$ & ... &  0.25  \\[2pt]
CL J0152.7-1357            & 0.830 & $0.45^{+0.17}_{-0.14}$ & $ 1.4^{+0.6}_{-0.5}$ & X--ray temperature & $16.96 \pm 0.08$ & $1.80 \pm 0.08$ &  0.23  \\[2pt]
XLSS J022303.0-043622      & 1.220 & $0.15^{+0.04}_{-0.03}$ & $0.68^{+0.20}_{-0.17}$ & X--ray temperature & $17.72 \pm 0.01$ & $1.82 \pm 0.01$ &  0.21  \\[2pt]
XLSS J022400.5-032526      & 0.810 & $0.19^{+0.05}_{-0.04}$ & $0.56^{+0.15}_{-0.13}$ & X--ray temperature & $16.49 \pm 0.10$ & ... &  0.34  \\[2pt]
RCS J0439-2904             & 0.950 & $0.03^{+0.01}_{-0.01}$ & $0.08^{+0.04}_{-0.02}$ & X--ray temperature & $17.70 \pm 0.08$ & $1.86 \pm 0.08$ &  0.14  \\[2pt]
2XMM J083026+524133        & 0.990 & $0.98^{+0.24}_{-0.21}$ & $ 4.0^{+1.1}_{-1.0}$ & X--ray temperature & $16.58 \pm 0.05$ & $1.90 \pm 0.06$ &  0.43  \\[2pt]
RX J0848.9+4452            & 1.260 & $0.47^{+0.17}_{-0.13}$ & $ 2.7^{+1.2}_{-0.8}$ & X--ray temperature & $17.00 \pm 0.02$ & $1.86$ &  0.42  \\[2pt]
RDCS J0910+5422            & 1.110 & $0.55^{+0.30}_{-0.19}$ & $ 2.7^{+1.8}_{-1.1}$ & X--ray temperature & $17.88 \pm 0.05$ & $1.83 \pm 0.06$ &  0.15  \\[2pt]
CL J1008.7+5342            & 0.870 & $0.19^{+0.10}_{-0.06}$ & $0.54^{+0.32}_{-0.19}$ & X--ray temperature & $16.42 \pm 0.08$ & $1.97 \pm 0.09$ &  0.41  \\[2pt]
RX J1053.7+5735 West       & 1.140 & $0.25^{+0.04}_{-0.03}$ & $ 1.1^{+0.2}_{-0.2}$ & X--ray temperature & $17.21 \pm 0.06$ & $1.99 \pm 0.07$ &  0.29  \\[2pt]
MS1054.4-0321             & 0.820 & $0.97^{+0.28}_{-0.22}$ & $ 3.4^{+1.1}_{-0.8}$ & X--ray temperature & $16.04 \pm 0.10$ & $1.80 \pm 0.10$ &  0.53  \\[2pt]
CL J1226+3332              & 0.890 & $ 1.8^{+0.4}_{-0.4}$ & $ 6.6^{+1.7}_{-1.5}$ & X--ray temperature & $16.00 \pm 0.06$ & $1.71 \pm 0.07$ &  0.63  \\[2pt]
RDCS J1252.9-2927          & 1.240 & $0.66^{+0.08}_{-0.11}$ & $ 3.9^{+0.6}_{-0.7}$ & X--ray temperature & $17.36 \pm 0.03$ & $1.83 \pm 0.01$ &  0.30  \\[2pt]
RDCS J1317+2911            & 0.810 & $0.24^{+0.19}_{-0.09}$ & $0.71^{+0.66}_{-0.29}$ & X--ray temperature & $17.27 \pm 0.15$ & $1.68 \pm 0.17$ &  0.17  \\[2pt]
WARPS J1415.1+3612         & 1.030 & $0.54^{+0.15}_{-0.12}$ & $ 2.3^{+0.8}_{-0.6}$ & X--ray temperature & $16.76 \pm 0.04$ & $1.86 \pm 0.05$ &  0.39  \\[2pt]
CL J1429.0+4241            & 0.920 & $0.57^{+0.32}_{-0.17}$ & $ 2.1^{+1.4}_{-0.7}$ & X--ray temperature & $17.43 \pm 0.20$ & $1.78 \pm 0.22$ &  0.18  \\[2pt]
CL J1559.1+6353            & 0.850 & $0.25^{+0.21}_{-0.11}$ & $0.74^{+0.72}_{-0.35}$ & X--ray temperature & $17.21 \pm 0.09$ & $1.90 \pm 0.12$ &  0.19  \\[2pt]
CL1604+4304               & 0.900 & $0.09^{+0.10}_{-0.04}$ & $0.23^{+0.30}_{-0.12}$ & X--ray temperature & $17.61 \pm 0.09$ & $1.68 \pm 0.12$ &  0.15  \\[2pt]
RCS J162009+2929.4         & 0.870 & $0.31^{+0.37}_{-0.14}$ & $0.95^{+1.33}_{-0.45}$ & X--ray temperature & $17.63 \pm 0.12$ & ... &  0.14  \\[2pt]
XMMXCS J2215.9-1738        & 1.460 & $0.18^{+0.06}_{-0.07}$ & $ 1.1^{+0.5}_{-0.5}$ & X--ray temperature & $18.72 \pm 0.01$ & $1.84 \pm 0.02$ &  0.11  \\[2pt]
XMMU J2235.3-2557          & 1.390 & $0.88^{+0.30}_{-0.24}$ & $ 6.5^{+2.7}_{-2.0}$ & X--ray temperature & $17.34 \pm 0.01$ & $1.87 \pm 0.02$ &  0.37  \\[2pt]

\hline
\end{tabular}
\end{table*}

\begin{table*}
\caption{BCGs without cluster mass measurements}\label{tab:nomass}
\centering           
\begin{tabular}{lrllr}
\hline
Name  & Redshift  &  Ks  &  J-Ks  &
Stellar Mass \\
     &  & [mag] & [mag] &    [$10^{12}\,M_{\odot}$] \\
\hline
Abell2292                 & 0.119 & $12.06$ & $1.06$ &  0.59  \\
Abell2665                 & 0.056 & $10.74$ & $1.11$ &  0.47  \\
Abell291                  & 0.196 & $14.10$ & $1.32$ &  0.22  \\
MACSJ0150.3-1005          & 0.363 & $13.90$ & ... &  0.82  \\
MACSJ0329.6-0211          & 0.451 & $14.13$ & ... &  1.07  \\
MACSJ1359.8+6231          & 0.330 & $14.32$ & ... &  0.45  \\
MACSJ2050.7+0123          & 0.333 & $14.67$ & ... &  0.33  \\
MACSJ2214.9-1359          & 0.495 & $14.71$ & $1.67$ &  0.76  \\
MACSJ2241.8+1732          & 0.317 & $14.39$ & ... &  0.39  \\
RCS0224-0002              & 0.770 & $16.87$ & ... &  0.22  \\

\hline
\end{tabular}
\end{table*}

\end{document}